\documentclass[preprint]{aastex}

\usepackage{bm}
\usepackage{graphicx}
\usepackage{rotating,makecell}
\usepackage{color}
\usepackage{amssymb}
\usepackage{lineno}
\usepackage{units}

\shorttitle{M-Dwarf HZ Climate Instabilities}
\shortauthors{Kite et al.}

\begin{document}
\linenumbers
%%\DeclareOption{nonatbib}


\title{Climate instability on tidally locked exoplanets}
%: Substellar weathering instability and substellar dissolution feedback.}
%\title{Warm-little-pond instability: atmospheric collapse on -locked planets}

%% any in the paper, not just in the front matter.

\author{Edwin S. Kite}%\altaffilmark{1}}
\affil{Department of Earth and Planetary Science, University of California at Berkeley, CA 94720}
\affil{Center for Integrative Planetary Science, University of California at Berkeley, CA 94720}
\email{kite@berkeley.edu}

\author{Eric Gaidos}
\affil{Department of Geology and Geophysics, University of Hawaii at Manoa, Honolulu, HI 96822}

\author{Michael Manga}%\altaffilmark{1}}
\affil{Department of Earth and Planetary Science, University of California at Berkeley, CA 94720}
\affil{Center for Integrative Planetary Science, University of California at Berkeley, CA 94720}

%\author{+ others, possibly, to be determined}


%\altaffiltext{1}{Center for Integrative Planetary Science, University of California, Berkeley}

\begin{abstract}
\noindent Feedbacks that can destabilize the climates of synchronously-rotating rocky planets may arise on planets with strong day-night surface temperature contrasts. Earth-like habitable planets maintain stable surface liquid water over geological time. This requires equilibrium between the temperature-dependent rate of greenhouse-gas consumption by weathering, and greenhouse-gas resupply by other processes. Detected small-radius exoplanets, and anticipated M-dwarf habitable-zone rocky planets, are expected to be in synchronous rotation (tidally locked). In this paper we investigate two hypothetical feedbacks that can destabilize climate on planets in synchronous rotation. (1) If small changes in pressure alter the temperature distribution across a planet's surface such that the weathering rate goes up when the pressure goes down, a runaway positive feedback occurs involving increasing weathering rate near the substellar point, decreasing pressure, and increasing substellar surface temperature. We call this feedback \emph{enhanced substellar weathering instability} (ESWI). (2) When decreases in pressure increase the fraction of surface area above the melting point (through reduced advective cooling of the substellar point), and the corresponding increase in volume of liquid causes net dissolution of the atmosphere, a further decrease in pressure will occur. This \emph{substellar dissolution feedback} (SDF) can also cause a runaway climate shift. We use an idealized energy balance model to map out the conditions under which these instabilities may occur. In this simplified model, the weathering runaway can shrink the habitable zone, and cause geologically rapid 10$^3$-fold atmospheric pressure shifts within the habitable zone. Mars may have undergone a weathering runaway in the past. 
Substellar dissolution is usually a negative feedback or weak positive feedback on changes in atmospheric pressure. It can only cause runaway changes for small, deep oceans and highly soluble atmospheric gases. Both instabilities are suppressed if the atmosphere has a high radiative efficiency. Our results are most relevant for atmospheres that are thin, have low greenhouse-gas radiative efficiency, and have a principal greenhouse gas that is also the main constituent of the atmosphere. These results identify a new pathway by which habitable-zone planets can undergo rapid climate shifts and become uninhabitable.
%The results from our simplified model motivate detailed Global Circulation Model (GCM) studies of the instabilities.  
\end{abstract}
%{\bf Talk about maximum entropy production. See what happens then.}
%also: {\bf Michael's comments: Make clear the seperation of timescales between atmospheric drawdown (slow) and the equilibration of surface temperature (fast).}
%also: We are assuming that the atmosphere is isothermal - that the atmosphere pays no heed to the surface temp. This is only OK for thin atmospheres, but thick atmospheres are invulnerable to the instability in any case (high $\chi$).




\section{Introduction}
%Gliese 581g, announcement of
%The very existence of a habitable zone is predicated upon the ubiquity of a negative climate feedback on ocean-bearing rocky planets.
%Here we propose a positive climate feedback ...
%Kepler has discovered $>$ 800 planet candidates. How many have maintained surface liquid water over geological time? 
\noindent Exoplanet research is driven in part by the hope of finding habitable planets beyond Earth \citep{exoptf}. Demonstrably habitable exoplanets maintain surface liquid water over geological time. Earth's long-term climate stability is believed to be maintained by a negative feedback between control of surface temperature by partial pressure of CO$_2$ (pCO$_2$), and temperature-dependent mineral weathering reactions that reduce pCO$_2$ \citep{hay81}. There is increasing evidence that this mechanism does, in fact, operate on Earth (\citet{coh04},\citet{zee08}, but see also \citet{edm03}). The Circumstellar Habitable Zone hypothesis \citep{kas93} extends this stabilizing feedback to rocky planets in general, between top-of-atmosphere flux limits set by the runaway greenhouse (upper limit) and condensation of thick CO$_2$ atmospheres (lower limit). H$_2$-H$_2$ collision-induced opacity can extend the habitable zone further out, in theory \citep{piegai,wordsbio}. Currently the best prospects for finding stable surface liquid water orbit M stars \citep{tar07,dem09}. Planets in M-dwarf habitable zones are close enough to their star for tidal despinning and synchronous rotation (\citet{mur99}, Chapter 5). Nearby M-dwarfs are the targets of several ongoing and proposed planet searches. Rocky exoplanets in hot orbits have recently been confirmed \citep{cor7b,kep10b,55cnce}, and are presumably in synchronous rotation. But does the habitable zone concept hold water for tidally locked planets? 

%Instabilities which threaten stable planetary climate are of three types. 
In this paper, we highlight two closely-related feedbacks which could cause climate destabilization on planets with and low-opacity atmospheres and atmospheres that do not have large optical depth. Both feedbacks require surface temperatures near the substellar point to be significantly higher than the planet-average surface temperature. % The dissolution and weathering instabilities can be much slower. However, provided there is some erosion to reveal fresh solids, all these instabilities occur in much less than the lifetime of a dwarf star.
%Atmospheric collapse on Early Mars occurs this way when at low obliquity when thick CO2 atmospheres cannot maintain the poles above the CO2 condensation point. 

%Of previously proposed feedbacks, these dayside instabilities are most closely related to 

\noindent %-\footnote{We assume 1:1 spin-orbit resonance throughout this paper}

%This hypothesis underpins instrument and mission design for space telescopes with a combined projected cost $>$\$5 bn(hyperbole?). 
%Destabilizing feedbacks have also been identified. Ice reflects sunlight, lowering temperature and promoting its own growth. Ice covered most of Earth 2.3 Ga and 0.7 Ga, and these extreme glaciations were in part due to ice-albedo feedback. (It is also currently assisting the Arctic Ocean's switch to a seasonally ice-free state).
 
%Atmospheric gases then dissolve in the liquid phase. 

\begin{itemize}
\item The \emph{enhanced substellar weathering instability} (ESWI) flows out of the same strong temperature dependence of silicate weathering that makes it possible for carbonate-silicate feedback to stabilize Earth's climate \citep{hay81}. Weathering and hence CO$_2$ drawdown rate increases rapidly with increasing temperature. Weathering also increases with rainfall, which increases with temperature \citep{ogo08,pie02,sch10}. Therefore, the global CO$_2$ loss rate depends heavily on the maximum temperature. On a synchronously-rotating planet where $\Delta T_s$ is high, most of the weathering occurs near the substellar point. Suppose weathering is initially adjusted to match net supply of greenhouse gases by other processes (e.g., volcanic degassing). Then consider a small increase in atmospheric pressure. Average temperature must increase, unless there is an antigreenhouse effect. Normally, this would lead to an increase in weathering. However, on a planet with a strong day-night temperature contrast, most of the weathering occurs near the substellar point. An increase in atmospheric pressure can decrease temperature around the substellar point, if increased advection of heat away from the hotspot by winds outweighs any increase in greenhouse forcing. Because this substellar area is cooling, and most of the weathering is around the substellar point, the planet-averaged weathering rate declines. Volcanic supply of greenhouse gases now outpaces removal by weathering, and a further increase in pressure occurs. This instability can lead to very strong greenhouse forcing, and may trigger a moist runaway greenhouse \citep{kas88}. Conversely, a small decrease in pressure from the unstable equilibrium can lead to atmospheric collapse. ESWI requires that weathering is an important sink for the major climate-controlling greenhouse gas, which is also the dominant atmospheric constituent. It also requires that the atmosphere is important in setting the mean surface temperature.  
%This can lead to very very high temperatures (or atmospheric collapse, for a small decrease in pressure).

\item \emph{Substellar dissolution feedback} (SDF) supposes an increasing gradient in surface temperature on an initially frozen planet, which allows a liquid phase to form (or be uncovered) around the substellar point. Some atmosphere dissolves in the new liquid phase. Positive feedback is possible if the decrease in atmospheric pressure due to dissolution raises the %planet's %maximum surface temperature. reas
temperature around the substellar point, increasing the fraction of the planet's surface area above the melting point. (We assume for the moment that $P$ exceeds the triple point). %atmosphere pressure further steepens the surface temperature gradient. 
In order for the mass of atmosphere sequestered in the pond to increase with decreasing pressure, increasing pond volume must outcompete both Henry's-law decrease in gas dissolved per unit volume, and the decrease in gas solubility with increasing temperature. For example, suppose $c \propto P$, where $c$ is the concentration of gas in the pond and $P$ is the atmospheric pressure, and $V \propto P^{-n}$ with $V$ the ocean volume. Then $n > 1$ is sufficient for positive feedback, and $n > 2$ is sufficient for runaway. %then the mass of atmosphere dissolved in the ocean will increase with decreasing pressure.
So long as the runaway condition is satisfied, the area of liquid stability will continue to expand: a pond becomes an ocean, drawing down the atmosphere. As with the ESWI, the key is that substellar temperature increases as pressure decreases. %This is more likely on  locked or slowly rotating planets, because thinner atmospheres will be less effective at transporting heat across the terminator to the nightside. 
Runaway SDF implies a climate bistability for a given inventory of volatile substance. One equilibrium has all of the volatile in the atmosphere. The other equilibrium has most of the volatile substance sequestered in a regional ocean and a little in the atmosphere, with the ocean prevented from completely freezing over by the steep temperature gradient that the thin atmosphere enables. A similar hysteresis was proposed for ancient Titan by \citet{lor97}. Runaway SDF is separate from the feedback between retreating ice cover and increased absorption of sunlight (ice-albedo feedback; \citet{roe10}), although both are likely to operate together. %Unlike the ESWI however, the existence of multiple equilibria is not sufficient to ensure a runaway instability.
\end{itemize}
%ESWI works with Planets with sunrise, not only 1:1 spin-orbit resonance, actually works for any planet with a strong day-night temperature contrast. However in this paper we focus on 1:1 because (discussion chunk).

\noindent Both instabilities occur more slowly than thermal equilibration of the atmosphere and surface. This separation of timescales allows us to solve for the fast processes that set the surface temperature (in \S2), and then separately address each of the two slower processes which may cause atmospheric pressure to change (in \S3-\S4).

Day-night color temperature contrast is among the `easiest' parameters to be measured for a transiting exoplanet \citep{cow11}, but there is currently no theory for $\Delta T_s$ on planets with observable surfaces. One motivation for this paper is to contribute to this emerging theory. We use a simplified approach which complements more sophisticated exoplanet Global Circulation Models (GCMs) \citep{jos97,jos03,mer10,eds11,wor11,pie11}. \S 2 sets out the energy balance model that is used for both instabilities, and \S 3 explains the enhanced substellar weathering instability including our choice of weathering parameterization. \S 4 explains the substellar dissolution feedback (considering only the 1:1 spin-orbit resonance). We find that SDF doesn't work in most cases, so readers motivated by short-term detectability can safely omit \S 4 and move to \S 5. \S 5.1 discusses relevant solar system data, including the possibility that Mars underwent a form of ESWI. \S 5.2 discusses applicability to exoplanets, and \S6 summarizes results. %, \S 3.2 discusses trigger mechanisms, and \S 3.3 a possible Solar System example (Mars). \S 4 summarizes results.
%The existence of our proposed feedbacks may be tested through our prediction of a bimodal distribution of day-night temperature contrasts on planets with surfaces.%

%\subsection{Heuristic explanation}

%\emph{Instability 1}

%
%\emph{Instability 2} - Warm little pond instability

\section{Idealized energy balance model}

\noindent Consider a planet in synchronous rotation on which surface liquid water is stable, with an atmospheric temperature that decreases with height along the dry adiabat. Slow rotation weakens the Coriolis effect, allowing the atmospheric circulation to all but eliminate horizontal gradients in atmospheric temperature at the top of the boundary layer, $T_a$. This is the weak temperature gradient approximation often made for Earth's tropics (e.g. \citet{mer10}). Figure \ref{GEOMETRY} shows the setup for our idealized energy balance model. The surface temperature $T_s(\psi)$ at an angular separation $\psi$ from the substellar point is set by the local surface energy balance:-

\begin{equation}
{SW_s(\psi)  - LW\!\!\uparrow\!\!(\psi) + LW\!\!\downarrow - \beta(T_{s}(\psi) - T_a) = 0}
\end{equation}

\noindent where $SW_s(\psi)$ is starlight absorbed by the surface,  $LW\!\!\!\uparrow\!\!(\psi) = \eta \sigma T_{s}(\psi)^4$ (where $\sigma$ is the Stefan-Boltzmann constant and $\eta \approx$ 1.0 is the emissivity at thermal wavelengths) is the surface thermal radiation, $LW\!\!\!\downarrow$ is backradiation from the atmosphere, $\beta$ is a turbulent heat transfer coefficient ($\beta = k_{TF} \,\,\rho$, where $\rho$ is the near-surface atmospheric density divided by Earth's sea-level atmospheric density, and $k_{TF}$ is a turbulent flux proportionality constant), and $T_a$ is the temperature of the atmosphere at the top of the boundary layer. (Equatorial superrotating jets can cause the hottest point on the surface to be downwind from the substellar point \citep{knu09,mit10,liu11}). The shortwave flux $SW_s(\psi) = L_* (1 - \alpha) \cos(\psi)$ corresponds to stellar flux $L_*$, attenuated by surface albedo $\alpha$. There is negligible transport of heat below the surface: we assume seas are not globally interconnected (or are shallow, or are deeply buried, or do not exist) and energy flux from the interior is small.

$LW\!\!\!\downarrow$ ( =  \( \frac{1}{2} \int_0^\pi \! LW\!\!\uparrow\!\!(\psi)\,  \sin\psi \, \mathrm{d}\psi  - OLR \), ignoring turbulent fluxes) % \textit{will need to fix this for final}) 
is longwave flux from the atmosphere to the surface. %We considered simple parameterizations for $OLR$, such as $\tau \propto \ln(P)$ where $\tau$ is optical thickness in the thermal. Though locally appropriate, these give grossly unrealistic results over the wide $P$ range of interest here. Instead, %
$OLR$ (Outgoing Longwave Radiation, longwave energy exiting the top of the atmosphere) is given by interpolation in a look-up table. To build this look-up table, we slightly modified R.T. Pierrehumbert's scripts at \texttt{http://geosci.\-uchicago.edu/$\sim$rtp1/Principles\- \\ PlanetaryClimate/}, particularly \texttt{PureCO2LR.py}. The look-up table gives $OLR$($P_{\Lambda},T_s$) for a pure noncondensing CO$_2$ atmosphere on the dry adiabat, with the temperature at the bottom of the adiabat equal to the energy-weighted average $T_s$, and with Earth gravity (9.8 m/s$^2$). Our noncondensing assumption introduces large errors for $T_s <$ 175K, so we assume that the top-of-atmosphere effective emissivity $OLR$/$LW\!\!\uparrow$ at these low temperatures is the same as at $T_s$ = 175K.

To investigate atmospheres not made of pure CO$_2$, we introduce an opacity ratio or relative radiative efficiency $\Lambda$, which is the ratio of the radiative efficiency of the atmosphere of interest to that of pure noncondensing CO$_2$. $\Lambda$ is a simplification of the complicated behavior of real gas mixtures \citep{pie10}. $\Lambda$ can be greater than 1 if the atmosphere contains radiatively very efficient gases (chloroflourocarbons, CH$_4$, NH$_3$, or the ``terraforming gases''; \citet{mar05}). We then query the look-up table using $P_{\Lambda} = \Lambda P$. Smaller values of $\Lambda$ have a weaker greenhouse effect (increased $OLR$).

%for an all-troposphere gray gas with a dry adiabatic lapse rate by (\citet{pie10}, eq. 4.32)

%\begin{equation}
%{OLR = \bar{I}_{+,s}\,\mathrm{exp}(-\tau) + \bar{I}_{+,s} \tau^{(4R/C_p)}\int_0^\tau \!  \tau_1^{(4R/C_p)} \mathrm{exp}(-\tau_1) \, \mathrm{d}\tau_1 }%+ I_{+,s}}
%\end{equation}

%$\bar{I}_{+,s} =  \frac{1}{4\pi} \int_0^\pi \!  LW\!\!\uparrow  A_{\psi} \, \mathrm{d}\psi$ is the weighted average longwave flux from the surface, $\tau$ is the optical depth of the atmosphere, $R$ is the gas constant, $C_p$ is the atmosphere's specific heat at constant pressure, and $\tau_1$ is a dummy optical depth increasing from zero at the top of the atmosphere to $\tau$ at the surface. The integral accounts for the decreasing probability that a photon emitted from within the atmosphere will escape to space as optical depth increases, and the exponent $(4R/C_p)$ relates temperature to optical depth on the dry adiabat. We use $R$ = 8.3145 J/K/mol and $C_p$ = 850 J/kg, appropriate for CO$_2$ gas at 300K. The corresponding $LW\!\!\downarrow$ is:-

%\begin{equation}
%{LW\!\!\downarrow =  \bar{I}_{+,s} \tau^{(4R/C_p)}\int_0^\tau \!  \tau_1^{(4R/C_p)} \mathrm{exp}(\tau-\tau_1) \, \mathrm{d}\tau_1}% + I_{+,s}}
%\end{equation}

Rayleigh scattering is relatively unimportant for planets orbiting M-dwarfs. Starlight is concentrated at red wavelengths where Rayleigh scattering is ineffective (falling off as $\lambda^{-4}$, where $\lambda$ is wavelength). The optical depth to Rayleigh scattering of 1 bar of Earth air is 0.16 for light from the Sun, but only 0.02 for light from the Super-Earth hosting M3 dwarf Gliese 581 (approximating both stars as blackbodies). In addition, absorption of starlight by the atmosphere is much stronger in the NIR than the visible, and so is more effective at compensating for Rayleigh scattering as star temperature decreases \citep{pie10}. We neglect Rayleigh scattering and absorption of starlight by the atmosphere. %increases with pressure and $\psi$. We take the Rayleigh scattering cross-sections for Earth air from Table 5.2 of \citet{pie10} -- other gases in oxidized atmospheres behave similarly (NH$_3$ and CH$_4$ scatter more; \citet{pie10}).

The horizontally uniform atmospheric boundary layer temperature, $T_a$, is set by the total energy balance of the atmosphere,

\begin{equation}
{\frac{1}{2} \int_0^\pi \!  \left[ LW\!\!\uparrow\!(\psi) + \beta(T_{s}(\psi) - T_a) \right]  \sin\psi \, \mathrm{d}\psi - OLR - LW\!\!\downarrow = 0}
\end{equation}

% $A_{\psi} = 2 \pi \sin\psi$ is a weighting factor for surface area,
\noindent where the integral gives the average flux from the surface. This reduces to $T_a = \nicefrac{1}{2} \int_0^\pi \! T_s(\psi) \sin\psi \, \mathrm{d}\psi$ because of our particular choice of $LW\!\!\downarrow$. In effect, we assume that the boundary layer only interacts with the ground through turbulent fluxes.

For a given $P_{\Lambda}$, we iterate to find $T_{a}$ and $T_{s}(\psi)$. $\psi$ resolution is 5 degrees. The initial condition has the surface in radiative equilibrium, and the atmosphere in equilibrium with this surface temperature distribution. Convergence tolerance is $\approx$ 2 $\times$ 10$^{-6}$.

%We let $\tau$ = $\Lambda ln(P + 1)$, %= \Lambda ln(k P + 1)$, 
%where $\Lambda$ is a specific radiative efficiency for the gas.

%k_vonk = 0.4; CpJkgK = 850; z1 = 10; z0 = 1e-4; %open water ... Pie10 p. 403.
%C_D = (k_vonk.^2)./(log(z1/z0).^2);
%kTF = CpJkgK*1.2*C_D*10 %cp*rho_s*C_D*U (for 1 bar and Earth gravity)
%kTF = 0.1738; %'tuned'

Throughout the paper, we assume $k_{TF} = C_p(T_a) C_D U$, where $C_p(T_a)$ is the temperature-dependent specific heat capacity of CO$_2$ ($\approx$850 J/kg/K at 300K), $C_D$ is a drag coefficient, and $U$ is characteristic near-surface wind speed. $C_D$ = $\frac{k_{VK}^2}{\ln(z_1/z_0)^2}$ where $k_{VK}$ = 0.4 is von Karman's constant, $z_1$ = 10m is a reference altitude, and $z_0$ is the surface roughness length ($10^{-4}$ m, which is bracketed by the measured values for sand, snow, and smooth mud flats; \citet{piel02}). For our reference $U$ = 10 m/s, this gives $k_{TF}\, \rho$ = 12.3 W/m$^2$/K. We assume a Prandtl number near unity. \S 5.2.1 reports sensitivity tests using different values of $U$.

We accept the following inconsistencies to reduce the complexity of the model:- (1) Radiative disequilibrium drives $T_s$ to a higher value than $T_a$ at the surface, and turbulence can never completely remove this difference. Therefore, setting the temperature at the bottom of the atmosphere to the surface temperature will lead to an overestimate of $LW\!\!\!\downarrow$ at low $P$. %The expression for $OLR$ assumes a gray gas, but we use a logarithmic dependence of $\tau(P)$. This is in line with real gas behaviour \citep{pie10} and prevents excessively high surface temperatures. 
(2) The expression for $k_{TF}$ is appropriate for a neutrally stable atmospheric surface layer, but turbulent mixing is inhibited on the nightside by a thermal inversion \citep{mer10}. This will tend to make the coupling between nightside atmosphere and nightside surface too strong. (3) We assume an all-troposphere atmosphere with horizontally uniform temperature. \citet{mer10} find temperatures are nearly horizontally uniform for Earthlike surface pressure and for levels in the atmosphere at pressures less than half the surface pressure. Atmospheric temperature is approximately horizontally uniform when the transit time for a parcel of gas across the nightside, $\tau_{advect} = \frac{a}{U_h}$, is short compared to the nightside radiative timescale, $\tau_{rad} \sim \frac{P}{g} \frac{C_p}{4 \epsilon \sigma T^3}$ \citep{sho10}. Here, $a$ is planet radius (1 Earth radius), $U_h$ is high-altitude wind speed ($\sim$ 30 m/s: \citet{mer10}), $\epsilon$ is an greenhouse parameter corresponding to the fraction of the emitted radiation that is not absorbed by the upper atmosphere and escapes to space, and $T$ = 250K is the atmospheric temperature, the radiative equilbrium temperature on the darkside being zero. Picking $\epsilon = 0.5$, this gives $\tau_{advect} \sim$ 2 days and $\tau_{rad} \sim$ 50 days. (4) The treatment of $T_a$ is crude. (5) We assume the atmosphere is transparent to stellar radiation, which is a crude approximation under M-dwarf (or cloudy) skies. (6) We neglect condensation within the atmosphere.

Representative temperature plots are shown in Figure \ref{TSE}, for $\Lambda$ = 0.1 and $L_*$ = 900 W/m$^2$. At low pressures, nightside temperatures are close to absolute zero, and substellar temperature is close to radiative equilibrium. Increasing pressure cools $\psi < 60^\circ$, and warms  $\psi > 70^\circ$. This is beause the atmosphere is warmer than the surface on the nightside, and cooler than the surface on the dayside. Therefore, the increase in $P (\propto \beta)$ increases the $\beta(T_s - T_a)$ term, which warms the nightside, but cools the dayside. For positive $\Lambda$, $LW\!\!\downarrow$ will increase with $P$ and warm the entire planet. However, for the relatively small value of $\Lambda$ shown here, the substellar point still undergoes net cooling with increasing pressure. This cooling with increasing $P$ is what makes the ESWI and SDF possible. When the surface becomes nearly isothermal, as for the ``10-bar'' curve in Figure \ref{TSE}, the entire surface warms with increasing pressure, and the ESWI and SDF cannot occur.

For $\Lambda$ $\ge$ 0, $\bar{T}_{s}$ must increase with $P$. Even if there is no greenhouse effect, the homogenization of the atmosphere will warm the planet on average because of the nonlinear dependence of $T$ on energy input \citep{eds11}. For small optical depth, nightside $T_s$ $\propto \ln(P)$. The fractional area of the planet over which liquid water is stable is 27\% at radiative equilibrium, decreasing with pressure and vanishing at $\sim$0.7 bars. As $P$ increases the greenhouse effect further, liquid stability reappears at $\sim$2.4 bars, rapidly becoming global. %The intervening pressure range in which liquid water is not stable anywhere on the planet is the ``dead zone,'' described for Mars by \citet{ric05}.

%We tuned the model against output of the LMD GCM for a 5\% CO$_2$, 95\% N$_2$ atmosphere for an Earth-sized planet (Robin Wordsworth, via email). Because our interest is in the dayside, we tune $k_{TF}$ and $\Lambda$ using the mean absolute deviation between dayside surface temperatures predicted by the simplified model and by the GCM.  The residual to the fit has mean absolute deviation 6K. This atmospheric composition gives best fit $k_{TF}$ = 0.43, $\Lambda$ = 0.3. %This is probably due to our neglect of absorption by sunlight in the atmosphere.

\section{Climate destabilization mechanism \#1: \\ Enhanced substellar weathering instability (ESWI)}

%\begin{equation}
%{\frac{W_{\psi}}{W_0} = \left( \frac{P}{P_0} \right)^{0.3} \mathrm{exp}\left(\frac{T_{s}(\psi) - T_{o}}{13.7} \right), T>T_{melt} }
%\end{equation}
\subsection{Weathering parameterisation}.

\noindent
The \citet{ber01} weathering relation, which is specific to CO$_2$ weathering of Ca-Mg silicate rocks, states

\begin{equation}
{\frac{W_{\psi}}{W_0} = \left( \frac{P}{P_0} \right)^{0.5} \underbrace{ \mathrm{exp}\left[k_{ACT}\,\,( T_{s}(\psi) - T_{o}) \right]}_{\rm direct\,\, T\,\, dependence} \underbrace{ \left[1 + k_{RUN}\,\,(T_{s}(\psi) - T_{o}) \right]^{0.65}}_{\rm hydrology \,\, dependence } }
\end{equation}

\noindent where $W_{\psi}$ is local weathering rate, $W_0$ is a reference weathering rate, $P$ is atmospheric pressure, $P_0$ is a reference pressure, $T_{o}$ = 273K is a reference temperature, $k_{ACT}$ = 0.09 is an activation energy coefficient, and $k_{RUN}$ is a temperature-runoff coefficient fit to Earth GCMs. Equation 3 is widely used, but uncertain and controversial (\S5.2.1). In our model, the strong temperature dependence leads to a strong concentration of weathering near the substellar point. For example, for a 1-bar atmosphere with $\Lambda$ = 0.1 (shown in Figure \ref{TSE}), 93\% of the weathering occurs in 10\% of the planet's area. The planet-averaged weathering rate is

%, and $T_{melt}$ is the melting point of water. 

\begin{equation}
{W_t(P) =  \frac{1}{4\pi} \int_0^\pi \! W_{\psi} A_{\psi} \, \mathrm{d}\psi}
\end{equation}

\noindent Climate is in equilibrium ($\frac{\partial P}{\partial t} $= 0) when planet-integrated weathering of greenhouse gases, $W_t$, is equal to net supply $V_n$ by other processes. %(These other processes can include volcanic degassing, metamorphic release, biology, sediment dissolution, and loss to space). We assume that the weathered greenhouse gas is also the principal gas in the atmosphere (see \S5.1 for discussion).  
The climate equilibrium is stable if $\mathrm{d} W_t / \mathrm{d} P > 0$ -- in this case, carbonate-silicate feedback enables long-term climate stability \citep{hay81}. Climate destabilization occurs when $\mathrm{d} W_t / \mathrm{d} P < 0$. The weathering feedback that underpins the Circumstellar Habitable Zone concept \citep{kas93} changes sign, and acts to destabilize these climates.

\subsection{ESWI results}
\noindent Figure \ref{WEATHERING} shows weathering rates corresponding to the temperatures in Figure \ref{TSE}. The lines show equilibria between weathering consumption of greenhouse gases and net supply by other processes (Figure \ref{WEATHERING}). The units of weathering are normalized to the weathering rate at $P$ = 3 mbar. Along the curves, planet-integrated weathering of greenhouse gases, $W_t$, is equal to net supply $V_n$ by other processes. These other processes can include volcanic degassing, metamorphic release, biology, sediment dissolution, and loss to space. The shape of this curve is set by competition between three effects:- (1) the greenhouse effect ($\Lambda$), (2) advective heat transport ($\beta, k_{TF}$) which enlarges and maintains unstable regions, and (3) stellar flux ($L_*$), which shrinks the unstable region. The curve has two stable branches and an unstable branch. The slope of the low-pressure stable branch is set by the $\left(P / P_o\right)$ term in Equation 5 - for small $P$, $\tau \propto \ln(P)$ and $\beta \propto P$ are small and the atmosphere has little effect on $T_s$. On the intermediate pressure unstable branch -- dashed in Figure \ref{TSE} -- the atmosphere is important to energy balance but $\beta$ outcompetes $\tau$. The homogenizing effect of $\beta$ cools the substellar region, and planet-integrated weathering decreases with increasing pressure. In the high-pressure stable branch, the planet is close to isothermal. Further increases in $\beta$ have little effect on $\Delta T_s$, but $\tau$ warms the whole planet and now is able to increase $W_t$. Catastrophic jumps in the pressure caused by small changes in the supply $V_n$ occur at $\sim$0.05 bar (for increasing volcanic activity) and $\sim$1 bars (for decreasing volcanic activity). The jumps correspond to a $>\!\!10^2$-fold increase in pressure, or a $>\!\!10^3$-fold decrease in pressure, respectively. The existence and location of these bifurcations are sensitive to small changes in the coefficients of (3). %For example, a change in the pressure exponent from 0.5 to 0.3 (the original value of \citet{wal81}) causes . 
The timescale for the climate regime jump is set by the rate of weathering and/or rate of volcanism on each specific planet. For example, Earth today supplies $\sim$15 mbar CO$_2$ in 10$^5$ yr (atmosphere + ocean: linearizing, $\sim$2 x 10$^7$ yr to build up 1 bar CO$_2$), but an Io-like rate of resurfacing \citep{rat04} with the same magmatic volatile content would build up 1 bar in $\sim$\emph{O}(10$^{4-5}$) yr. A natural weathering-rate experiment occurred on Earth 0.054 Gya, with very rapid release of CO$_2$ from an unknown source. The warmed climate required $\Delta t$ $\sim$ \emph{O}(10$^5$) yr (\citet{mur10}, using $^3$He accumulation dating) to draw down 0.9 mbar of excess CO$_2$ \citep{zee09}. 

Figure \ref{LSTARP} shows habitable-zone climate regimes as a function of equilibrium pressure and stellar flux, which can change (Figure \ref{LSTARP}a) due to stellar evolution, tidal migration \citep{jac10}, or close encounters with other planets and small bodies \citep{nicerecent}. Climate stability depends strongly on $\Lambda$, so we show this for three values of $\Lambda$ - an almost radiatively inert gas (Figure \ref{LSTARP}b), an intermediate $\Lambda$ = 0.1 case (Figure \ref{LSTARP}c), and a strong opacity that only just allows the ESWI (Figure \ref{LSTARP}d, for $\Lambda$ = 1.0; higher values are stable against ESWI). %Stability of weathering equilibria $W_t = V_n$ is shown by thin solid contours -- thin cool-colored lines are unstable equilibria $\frac{\partial W}{\partial P} <$ 0, thin warm-colored lines are stable equilibria $\frac{\partial W}{\partial P} >$ 0. 
The thick black line labelled with zeros corresponds to marginally stable climate equilibrium. Increasing $L_*$ widens the range of $P$ that sits within the low-pressure stable branch. That is because higher $L_*$ produces higher absolute temperatures, and higher absolute temperatures favor radiative exchange between atmosphere and surface ($\propto$ 4$(T_s - T_a)^3$ for small $(T_s - T_a)$ ) versus turbulent exchange ($\propto (T_s - T_a)^1$). Therefore, a given equilibrium value of $P$ on the low pressure branch is more stable at low $L_*$ than high $L_*$ :-- for a small increase in $P$ the radiative warming will be less counteracted by the cooling of the substellar point. Increasing $L_*$ pushes the the high pressure branch to higher values. %\emph{Edwin notes:  I don't know why}.%(We allow weathering to take place for $T < T_{melt}$, but we see the same qualitative behaviour if we set weathering to zero for $T < T_{melt}$.																																																								
Other instabilities that contain the habitable zone are shown by thin lines. The dash-dotted line corresponds to pressures in excess of the nightside CO$_2$ saturation vapor pressure. CO$_2$ atmospheres to the left of this line condense on fast, dynamical timescales. Increasing $\Lambda$ couples the horizontally-isothermal atmosphere more strongly to the nightside surface, and causes the CO$_2$ collapse threshhold to move to the left. %Climates to the left of the thin dotted line correspond to entirely frozen Snowball planets . Increasing $\Lambda$ raises substellar temperature at high $P$, so this line also moves left with increasing $\Lambda$. 
The thin solid line corresponds to mean surface temperatures above 40$^\circ$C, which is the lower edge of the moist runaway greenhouse zone \citep{kas88,pie95} - our model does not include fluxes of latent heat or the greenhouse effect of water vapor, so the position of this line is notional.
The maximum pressure and minimum pressure of the bifurcation loop (Figure \ref{WEATHERING}) are shown by thick gray lines in figure \ref{LSTARP}. In some cases, the lower end-of-transition pressures are so low that water boils (thin dashed line). This would initiate the boiling of a global ocean. If boiling continued, the eventual fate could be a steam ocean, or restriction of liquid water stability to a thin belt near the terminator. The upper end-of-transition pressures are often $>$10 bars, with a nearly isothermal $T_s$. Such planets would have weak and perhaps undetectable phase curves in the absorbing bands \citep{sel11}.

For the almost radiatively inert gas (Figure \ref{LSTARP}b), geologically stable equilibria usually have $P<$0.01 bars or $P>$0.3 bar. Intermediate pressures cannot be stable over geological time. The overall pattern is similar for $\Lambda$ = 0.3 (Figure \ref{LSTARP}c). Increasing $\Lambda$ always shrinks the domain of the unstable branch. For much higher $\Lambda$ the climate is stable everywhere. We show the last gasp of the instability in Figure \ref{LSTARP}d. Higher radiative efficiency means that for a small change in $P$, when $\Delta T_s$ is significant, radiative heat transfer
overcomes advective heat transfer, the substellar patch warms, and overall planet temperature increases.

Figure \ref{HYSTERESIS} summarizes the effects of ESWI in a stability phase diagram (against axes of gas radiative efficiency and incident
stellar radiation). For $\Lambda$ $>$ $\sim$1, the climate is stable to ESWI. For $\Lambda$ $<$ $\sim$1, ESWI is possible but the pressure jumps caused by ESWI do not always have a catastrophic effect. Higher $L_*$ warms the climate towards the runaway moist greenhouse threshhold, and upward jumps in pressure for $L_* >$ $\sim$2000 W m$^{-2}$ may initiate the moist runaway greenhouse (points to the right of the vertical dashed line in Figure 5). % (the 5\% CO$_2$, 95\% N$_2$ atmosphere we tuned to is apparently stable). 
%The runaway greenhouse is only possible when ; below this value, insolation is insufficient to bring the mean temperature above the greenhouse threshhold. 
Atmospheric collapse to $\sim$ 1 mbar (well below the triple point of water) only occurs below $\Lambda$ $<$ 0.4. % because \emph{Edwin notes: I have no idea why.}.
Increasing $L_*$ increases both the bottom and the top pressure for instability, implying that tidal migration towards the star should be destabilizing for thick atmospheres but stabilizing for thin atmospheres. %However, this ignores the tidal effect on volcanism, which -- for fixed eccentricity -- should increase as the planet spirals in \citep{beh11}. Increasing rates of volcanic degassing increase the weathering rate in equilibrium with greenhouse gas supply. Within either stable branch, this means the climates of inspiralling planets move up in Figure \ref{LSTARP} at the same time as they move to the right. In summary, the ESWI does create a large swath of climate instability within the habitable zone, but only for small values of $\Lambda$.
%(Escape to space should be a small effect for tectonically active planets within the habitable zone \citep{tia09} \emph{Eric, is this correct?}).

%Note on doesn't work for Earth/

%All known planetary surfaces are made of either water ice or basalt.\footnote{Excluding polar caps, superficial coatings, and plagioclase-rich scum.} 

\section{Climate destabilization mechanism \#2: Substellar dissolution feedback (SDF)}
\noindent Water ice and basalt are the most common planetary surface materials in the Solar System, and are expected to be common elsewhere.When these melt (around 273K for ice and 1300K for basalt), atmosphere can dissolve into the melt.
%\emph{1. Water ice: Dissolution-assisted snowball deglaciation.} \\
%\emph{2. Basalt: Steam atmosphere collapse into magma pond.}
Counterintuitively, a decrease in $P$ and in average surface temperature $\bar{T}_s$ can favor melting if $\Delta T_s$ is large, as pointed out for Mars by \citet{ric05}.
For a synchronously-rotating planet entirely coated in condensed material (ice, rock, or carbon-rich ceramic), surface liquid will first appear near the substellar point. Atmosphere will dissolve into this warm little pond, approaching Henry's-law equilibrium:

\begin{equation}
{P_{pond} = \frac{g}{P_{Earth}} \underbrace{ \left( D_{pond} \,\frac{1}{2} \left(1 - \cos\psi_{max} \right) \right) }_{\rm global-equivalent\,\, liquid\,\, depth} \, \underbrace{ \left(\,P^d\, m\,  \rho_l \, k_{H}(T^o)\, \mathrm{exp}\!\left[ C \left(\frac{1}{T_{pond}} - \frac{1}{T^o} \right) \right] \right)}_{\rm mass \,\, of \,\, gas \,\,per\,\, unit\,\, liquid \,\, volume} }
\end{equation}

\noindent where $P_{pond}$ (in bars) is the equivalent atmospheric pressure of gases dissolved in the ocean, $g$ is surface gravity, $P_{Earth}$ = $1.01 \,\, \mathrm{x} \,\, 10^5$ Pa is a normalization constant, $D_{pond}$ is pond depth, $\rho_l$ is liquid density, $\psi_{max}$ is the angular radius of the pond,  $P$ is the surface pressure, $d$ is a dissolution exponent ($\sim$0.5 for water in silicate liquids and $\sim$1 for gases in water),  $m$ is the molecular mass of the atmosphere, $k_H$ and $C$ are Henry's-law coefficients, $T_{pond}$ is the pond temperature, and $T^o$ is a reference temperature. Here, the first term in brackets is the depth of the pond in global-equivalent meters, and the second term in brackets is the mass of gas per unit volume of pond. We have neglected the distinction between fugacity and partial pressure. The pond is assumed to be well-mixed so that $T_{pond}$ = $ \frac{1}{1 - \cos\psi_{max}} \int_0^{\psi_{max}} \! T_{s,\psi} \sin \psi \, \mathrm{d}\psi$. This relation assumes a uniform heat transfer coefficient between the surface and the pond.

%Per unit volume, Henry's Law will . However
%The pond will grow to become an ocean, so long as
Instability occurs when a
decrease (increase) in surface pressure results in an uptake (release)
of gases from the pond that exceeds that which is consistent with that
change in pressure, i.e. \( \frac{\partial P_{pond}}{\partial P} < -1\). In this case the feedback has infinite gain \citep{roe09}.
%Pond growth is a positive feedback on climate change if the loss of atmosphere by dissolution leads to an expansion of the zone where liquid is stable, and %any exsolution associated with increasing ocean temperature is insufficient to decrease the 
%total dissolved mass in the ocean increases with decreasing pressure. Runaway instability occurs when a feedback has infinite gain \citep{roe09}. For SDF, this corresponds 
%\( \frac{\partial P_{pond} }{\partial P} < -1 \).
%For low starting pressures, pressure can be drawn down to the triple point pressure. Only for rare cases does temperature dependence outweigh the ocean volume increase.
Pond growth rate is limited by the balance between insolation and the latent heat of melting (\emph{O}(10$^3$) yr for melting of a 1km-thick ice sheet, Earthlike insolation, $\alpha$ = 0.6, and 10\% of sunlight going to melting), or by thermal diffusion of heat into a stratified pond. For tectonic activity levels that are not too high, the SDF is much faster than the ESWI, which is limited by volcanic degassing rates and tectonic resurfacing rates (0.05 mm/yr on Earth \citep{lee99}; 16 mm/yr on Io \citep{rat04}). The SDF stops when insolation is insufficient to allow further pond growth. What happens after the SDF has operated will depend on the sign of the carbonate-silicate feedback at the new, modifed $P$. If $\partial W_t / \partial P$ $>$ 0, the normal carbonate-silicate feedback will rejuvenate the atmosphere on a volcanic degassing timescale, freezing the ocean. If $\partial W_t / \partial P$ $<$ 0, pressure decreases further.

%(A second, more restricted SDF branch occurs for complete ocean coverage and increasing pressure. 
%Then temp goes up, solubility goes down.)

Gases that react chemically with seawater (such as SO$_2$ and CO$_2$; \citet{zee01}) can have $P_{pond}$ much greater than that given by Henry's law. For example, total Dissolved Inorganic Carbon (DIC, $\propto P_{pond}$) is buffered against changes in $P$ by carbonate chemistry, and $P_{pond}$ changes much more slowly than Henry's law. $P_{pond} \propto P^{0.1}$ for the modern Earth ocean, as opposed to $P_{pond} \propto P$ for Henry's law \citep{zee01,goo09}. For a decrease in $P$ that increases $\psi_{max}$, this buffering favors the tendency of increasing pond volume to draw down more atmosphere, against the Henry's Law decrease in atmospheric concentration per unit pond. Carbonate buffering is less important for $P > \sim$1 bar, because at the correspondingly low pH the DIC partitions almost entirely into CO$_2$. For CO$_2$, we use R. Zeebe's scripts (\texttt{http://www.soest.hawaii.edu/oceanography/faculty/zeebe\_files/CO2\_System\_in\_Seawater /csys.html}) to find the additional DIC held in the ocean as HCO$_3^-$ and CO$_3^{2-}$. We use fixed alkalinity, 2400 $\mu$mol/kg (similar to the present Earth ocean; \citet{zee01}). Figure \ref{SDICO2} shows the results. $\psi_{max}$ is set by Equation 1 through $T_{s}(\psi)$. Ocean circulation adds additional heat transport terms to Equation 1, which we ignore. We also do not consider buffering by dissolution and precipitation of carbonates or salts (such as sulfates).

As an example of SDF, consider the partitioning of an initial 1-bar total inventory of CO$_2$ (blue-green contour labelled `0' in Figure \ref{SDICO2}) with increasing $L_*$. Initially, the planet's surface is below freezing. As stellar flux is increased to 800 W/m$^2$, a small sea forms and dissolves some of the CO$_2$. The system is within the area where SDF is a positive feedback on small changes in $P$ (outermost black contour in Figure \ref{SDICO2}), which accelerates sea growth. A small further increase in $L_*$ leads to runaway SDF (innermost black contour in Figure \ref{SDICO2}), and ocean area quickly grows with from $\sim$5\% to $\sim$15\% of planet surface area with no change in $L_*$. After this change the atmospheric inventory is reduced to \nicefrac{1}{4} bar with the remaining \nicefrac{3}{4} bar dissolved in the ocean. Further pond growth requires further increase in $L_*$. Increasing $T_{pond}$ decreases solubility and shallows the slope of increasing volatile inventory stored in the ocean. By $L_*$ = 3800 W/m$^2$ (the highest considered), the ocean covers almost the entire lightside hemisphere, and stores $\sim$\nicefrac{5}{6} of the initial CO$_2$ inventory. Further small increases in $L_*$ cause a global ocean.

Three main effects control this system:- (1) Cosine fall-off of starlight weakens the ability of decreasing pressure to increase ocean area beyond a relatively small $\psi_{max}$. Following a line of decreasing pressure, the dashed red lines (fractional ocean area) become more widely spaced with decreasing pressure. Cosine falloff of stellar radiation is responsible. This restricts the scope of the instability, which tends to lead to Eyeball states \citep{pie11}. We do not find any cases where the SDF can turn a dry planet into an ocean-covered planet or vice versa. (2) Ocean instability disappears above $\sim$5 bars, when the surface is nearly isothermal. Because decreases in pressure always decrease $\bar{T}_s$ (except when there is an antigreenhouse), decreases in pressure can only cause oceans to freeze over. Below $\sim$5 bars, $\Delta T_s$ is not negligible. If a decrease in pressure allows the substellar temperature to rise above freezing, an ocean can form. (3) Rectification of starlight by the terminator, and of ocean area by the melting point, divides the phase diagram into `no ocean', `substellar ocean' and `global ocean' zones. On the nightside $T_s$ is constant, so ocean area jumps from 50\% to 100\%. Notice the change in sign of $L_*$ dependence below and above the dashed red line corresponding to 50\% ocean area. In the substellar ocean zone, increasing $L_*$ increases ocean area and the ocean inventory increases. However, in the global ocean zone, increasing $L_*$ cannot increase ocean area. The decrease in gas solubility with increasing temperature dominates, and the ocean inventory decreases.

Figure \ref{SDICO2} does show unstable regions in parameter space for which substellar dissolution is a positive feedback on changes in $P$ (solid black lines). These always correspond to small oceans ($<$10\% of planet surface area). But %, except in the high-pressure limit where the planet `jumps' from no ocean to a near-global ocean. 
the gain of the feedback is small, and runaways cannot occur unless ocean depth $>$10km. %Relatively strong positive feedback requires $\Lambda <$ 0.1, but this may not be relevant for thick CO$_2$-rich atmospheres.% because the best fit to a GCM run at 5\%CO$_2$, 95\% N$_2$ is $\Lambda \sim$ 0.3. 
Factor--of--3 decreases (or increases) in atmospheric pressure can occur with no change in the total (atmosphere + ocean) inventory of volatile substance. %For pressures $>$ 3 bars in this case, SDF cannot occur because the thick atmosphere homogenizes surface temperature. As a result, small decreases in pressure tend to decrease pond volume, not increase it. 
The SDF is a minor positive feedback (or a negative feedback) for most gases which dissolve simply according to Henry's Law. The buffering effect is not sufficient to allow ocean area to give a strong positive feedback. Finally, $\Lambda$ = 0.03 is unrealistically low for all-CO$_2$ atmospheres. Setting $\Lambda$ = 1 would shut down the instability. We conclude that the CO$_2$ SDF is unlikely to be important for planets in synchronous rotation, and is important in fewer cases than is the ice-albedo feedback \citep{roe10,pieareps}. %\emph{NOT YET DONE! We looked at SO$_2$ \citep{abd77} and reached the conclusion that X. NOT YET DONE!}

%\emph{SO$_2$? \citep{abd77} As a limiting case, we consider SO$_2$, which is highly soluble in water ($k_{H}(T^o)$ = 1.2 M/bar and $C$ = 3100 K, \citet{sancomp}), and we assume $D_{pond}$ = 5km. Our approach is to calculate gas dissolved in ocean as a function of atmospheric pressure. In most cases, we find ${\frac{\partial P_{pond} }{\partial P} > 0}$ . However, for small fractional ocean area $< 0.15$, and correspondingly low pool $T_{pool} <$ 285K, we find that the feedback can draw down pressure by a factor of 20, through rapid growth of the pond.}

%More GCM points are desirable.

%(Discuss the figure).

%We ignore ice-albedo feedbacks, but these will only amplify this instability \citep{pie11}.% (Maybe go into more detail on ice albedo feedback?)
%We also ignore nonlinear feedbacks between $T_s$ and $T_a$.

%In the absence of ice albedo feedback, the pond dissolution feedback has only a small gain (stabilizing or mildly destabilizing; \citet{appropriate roe}) and does not run away . Henry's Law defeats the increase of 

%(%This is for cold planets.)

%What kind of perturbations can trigger? cont drift, volcanic plume, tidal pulse volcanic effect, tidal pulse delta A effect ...

%Need more GCM points for calibration.

\section{Discussion}
\noindent

%\subsection{Earth and Solar System Data}
\subsection{Solar system climate stability - is Mars a solar system example of ESWI?}

% \emph{Snowball Earth collapse} via ice-albedo feedback \citep{roe10} can occur without atmospheric feedback, but is also limited by the need to radiate  latent heat.
%Earth's steam ocean most likely ended this way \citep{zahnle}. Atmospheric collapses are rate-limited by the surface's ability to radiate the heat of condensation.  reducing both the greenhouse effect and the dampening of the meridional temperature gradient }.  

\noindent Although ESWI is most relevant for planets in synchronous rotation, it can work for any planet with sufficiently high $\Delta T_s$. Therefore, in order to test ESWI against available data, we compare the requirements for ESWI to Solar System data (Table 1). After correcting for the distorting effects of life, all of the Solar System's non-giant atmospheres are overwhelmingly one gas. Except for Earth, the principal gas is also the main greenhouse gas. %Not if they all weather or 
Venus' atmospheric composition is not controlled by the abundance of surface liquid (nor in solid-state equilibrium with surface minerals; \citet{has05,tre11}), and Triton's atmosphere is too thin to stabilize liquid nitrogen. Climate regulation on Titan is not well understood. Currently, the greenhouse effect of CH$_4$ outcompetes the antigreenhouse effect of the organic haze layer \citep{mck91}. The production rate of the organic haze layer depends on [CH$_4$] \citep{mck91,lor97}. ESWI is not currently possible on Titan because $\Delta T_s$ is too small, 2.5-3.5K \citep{jen09}. Therefore, out of 5 nearby worlds with atmospheres and surfaces, only Mars is a candidate for ESWI (\S5.1). Solar System data suggest that the conditions for the ESWI are quite restrictive, and that most planets will not be susceptible to the ESWI.
%dissolve at the same rate .%
%Problem is definition of equilibrium because the degassing will in general be different. 

Mars may have passed through ESWI in the past. The main climate-controlling greenhouse gas (CO$_2$) can dissolve in liquid water and be sequestered as carbonate minerals. Mars has low $P$ (95\% CO$_2$). It has widespread deposits of carbonate minerals \citep{ban03,wra11}, but little or no surface liquid water.  It currently sits at the gas-liquid sublimation boundary \citep{kah85}, with $\Delta T_s \sim$100K , and GCMs show that $\Delta T_s\!\!\downarrow$ as $P\!\!\uparrow$ \citep{ric05}. These geologic and climatic observations are all consistent with a past rapid transition via ESWI from an early thicker atmosphere to the current state, as follows. Increasing solar luminosity could have permitted transient liquid water, allowing carbonate formation. The corresponding drawdown in $P$ would increase noontime temperature, allowing a further increase in liquid water availability and the rate of carbonate formation. The runaway would slow as $P$ approached the triple-point buffer: loss of liquid water stability may have throttled weathering and buffered the climate near the triple point \citep{kah85,hal07}. However, ice is not ubiquitous on the surface and can migrate away from warm spots, so unusual orbital conditions are neccessary for melting \citep{kit11a,kit11b}. Alongside carbonate formation, atmospheric escape, polar cold trapping and volcanic degassing are the four main processes affecting $P$ on Mars over the last 3 Ga. However, recent volcanism has been sluggish and probably CO$_2$-poor \citep{hir08,wer09,sta11}, and present-day atmospheric escape appears slow \citep{bar07}. Polar cold traps hold $\sim$1 Mars atmosphere of CO$_2$ as ice today \citep{phi11}, but this trapping should be reversable at high obliquity.
Therefore, it is quite possible that substellar ($\equiv$ noontime) carbonate formation has been the dominant flux affecting the secular evolution of Mars' atmosphere since 3.0 Gya. This hypothesis deserves further investigation.

%\subsubsection{Solar System climate instabilities}
%\noindent Previous work on Solar System objects has identified three different timescales ($\Delta t$) for destabilization.  Timescale 1 ($\Delta t$ $\sim$ 10$^{0-1}$ yr): instabilities rate-limited by atmospheric dynamics. These include \emph{atmospheric collapses} such as the seasonally reversing partial atmospheric collapse that occurs today at Mars' winter pole \citep{hab94,rea04}; photochemical collapses, which may have occurred on Mars \citep{zah08} and Titan \citep{lor97}; and greenhouse runaways \citep{kas88,lor99,sug02}. Type 2 ($\Delta t$ $\sim$ 10$^3$ yr) instabilities are paced by the mixing of a stratified or slowly-overturning fluid reservoir. This includes Earth's ocean thermohaline circulation bistability \citep{sto61,epi06} . %In a stratified ocean this requires molecular diffusion across internal boundary layers, $\Delta t$ $\sim$ \emph{O}(10$^6$) yr. 
%Type 3 ($\Delta t$ $\sim$ \emph{O}(10$^{5-9}$) yr) instabilities are rate-limited by the availability of fresh surface area for weathering. For example, rebalancing Earth's climate following a temperature spike 0.054 Gya required  $\Delta t$ $\sim$ \emph{O}(10$^5$) yr (\citet{mur10}, using $^3$He accumulation dating) to draw down 0.9 mbar CO$_2$ \citep{zee09}. Volcanic/tectonic resurfacing rate is very sensitive to mantle composition, tidal heating, and tectonic style \citep{kit09,val09,kor10,beh10,van11}, explaining the wide range in $\Delta t$. 

%\noindent ESWI is of type 3. SDF is of type 2.

\subsection{Application to exoplanets}

\subsubsection{How general is our feedback?}

%\emph{\underline{Incorporate sensitivity tests here.}}

\noindent ESWI requires: 

\noindent -- \emph{Strong temperature dependence of the weathering drawdown of greenhouse gases}. Decreasing $k_{ACT}$ to 0.03 (from the nominal 0.09) eliminates the ESWI except for radiatively inert atmospheric compositions. On the other hand, increasing $k_{ACT}$ to 0.27 causes a large unstable region even for $\Lambda$ = 1.0, with at least a 1 dex range of atmospheric pressure unstable to ESWI for all habitable-zone luminosities. 

Earth data on the value of $k_{ACT}$ are consistent with $k_{ACT}$ $\sim$ 0.1. However, deep-time calibration of weathering-temperature relations such as Equation 3 is difficult. There is only one planet (Earth) to use as laboratory, with constantly drifting boundary conditions, and rather few natural experiments. Chemical weathering rates of silicate minerals in the lab are definitely temperature-dependent \citep{whi95}, but erosion-rate dependence is also important at the scale of river catchments \citep{wes05}. Confirming temperature dependence on geological scales is difficult, in part because today's weathering rate contains echoes of glacial-interglacial cycles \citep{van09}.  Regression of present-day river loads on watershed climatology by \citet{wes05} suggests an \emph{e}-folding temperature of 8.5(+5.5/-2.9)K. $^{187}$Os/$^{188}$Os data suggest continental weathering rates increased 4-8x in $\ll$10$^{6}$ yr during a Jurassic hyperthermal ($\Delta T$ $\le$ 10K) 0.183 Gya \citep{coh04}, implying an \emph{e}-folding temperature $<$(5 -- 7)K. Analysis of the apparent time dependence of weathering rate gives support to a hydrological control on weathering rates \citep{mah10}, but on a planetary scale precipitation always increases when $T_s$ increases \citep{ogo08}. That is, \( \frac{D W}{D t} = \frac{\partial W}{\partial T_s} + \frac{\partial W}{\partial R}\frac{\partial R}{\partial T_s} \) with \( \frac{\partial R}{\partial T_s} > 0 \).

%.... Laboratory measurements show \citep{areps00} ....

%Recovery timescale for an even more severe hyperthermal ($\Delta T$ up to 5-9 K) 0.056 Gya suggests weathering. 
%Rainfall increases with temperature \citep{ogo08}, which also favors weathering.% and which we fold into the temperature dependence. 
%The \citet{hay81} relation, which is specific to CO2 weathering of silicate rocks,

Overall, Equation 3 is consistent with deep-time, present-epoch, and laboratory estimates for Earth. Though Equation 3 is used in this paper as a general rule for Earth-like planets, the weathering-temperature relation is shaped by biological innovations. For example, the symbiosis between vascular plants and root fungi (arbuscular mycorrhizae) acidifies soil, profoundly accelerates weathering, and may be unique to Earth \citep{tay09}. All geologically important surface weathering reactions require a liquid phase \citep{whi95}. We assume weathering reactions do continue below the freezing point (due to microclimates, or monolayers of water). However, the results shown in Figure \ref{LSTARP} and \ref{HYSTERESIS} did not change qualitatively when we set $W = 0$ for $T < T_{melt}$.
%Here a few sensitivity tests \emph{motivated by solar system data}. In one, we allow wind speed to decrease with increasing atmos pressure. 

%In another, we crank down the P sensitivity in the weahtering relation \citep{tay09.}

Strong temperature dependence could break down in several ways. For example, erosion is needed to expose fresh mineral surfaces for weathering. On a tectonically quiescent planet (and for tectonically quiescent regions of an active planet) weathering may be limited by supply of fresh surfaces, with weathering going to completion on all exposed silicate minerals. On tectonically very active planets with little or no land, hydrothermal alteration and rapid seafloor spreading maintain low greenhouse gas levels with little or no temperature dependence \citep{sle01}. A planet with weathering rates that are limited by the availability of liquid water (like Mars) can have effectively temperature-dependent weathering, but only if the atmospheric pressure is well above that fluid's triple point.

\noindent -- \emph{Large $\Delta T_s$}. The high-$\Delta T_s$ requirement cannot be met if the atmosphere is thick. A deep global ocean circulation behaves like a thick atmosphere - Earth abyssal temperatures vary $<$2K from tropics to poles \citep{sch00}. Therefore, the climate destabilization mechanism cannot operate on a planet with a deep global ocean circulation - at least some land is necessary.  Evaporation will dry out land at the substellar point if $\bar{T}_s$ is high, so weathering activity may be concentrated at cooler $\psi$ in this case.  $\Delta T_s$ varies little, or even increases, with rotation frequency \citep{mer10,eds11}. Therefore, ESWI could work for rapidly-rotating planets such as Mars (\S 5.1). However, the isothermal approximation does not apply when the Coriolis force prevents fast equator-to-pole winds. For rapid rotators, $T_s$ is a function of latitude and longitude, and our idealized energy balance model is not appropriate. %, not just distance from the substellar point.

%Weathering skin depth (the depth range of material subject to the highest temperatures, water activity, and so on) decreases for planets not in 1:1 spin-orbit resonance. On a planet in 1:1 spin-orbit resonance this is the whole crust, \emph{O}(10$^{4}$ m). If spin is rapid and thermal diffusivity is low this can drop to \emph{O}(10$^{-2}$ m). This would make the rate of climate system response to perturbations more sensitive to resurfacing rates. Water can drip deeper than the thermal skin depth.

%Faster rotation rate would be a big deal for low pressure atmospheres, and mainly affect the minimum temperature so chances of dynamical collapse of atmosphere.
\noindent -- \emph{Small $\Lambda$}. Strong greenhouse gases have high $\Lambda$, which suppresses ESWI. On the other hand, $\Lambda$ can be negative if there is an antigreenhouse effect ($\Lambda$ $<$ 0). M-dwarfs later than M4, with fully convective interiors, seem to remain active with high UV fluxes for much longer than do Sunlike stars.
High UV fluxes broadly favor CH$_4$ accumulation \citep{seg05} and perhaps antigreenhouse haze effects. When $\Lambda <$ 0, ESWI will apply for all $P$ and $L_*$. 
%ESWI could be unimportant on antigreenhouse planets if surface weathering is a small sink for the atmosphere compared to soot formation. In that case, the classical weathering feedback \citep{hay81} cannot stabilize the atmosphere either. If the source of CH$_4$ is biology rather than geological activity, then feedbacks involving an optimal temperature for methanogens may be important (e.g., \citet{dom08}).

%We assume 1:1 but weathering instability should apply for any planet for which TIU convert to a time is much less than rotation frequency $\Omega$.
%\begin{equation}
 %x = y;
%\end{equation}
%(Example: rock ice)

\noindent -- \emph{Surface-atmosphere coupling}. This consists of radiative and turbulent coupling. Turbulent coupling requires a nonzero near-surface wind speed, and that the global near-surface atmosphere is not stably stratified. A sensitivity test with a 10--fold reduction in $U$ showed that the pressure range unstable to ESWI moves to $\sim$ 10$\times$ higher pressure. The range of $\Lambda$ subject to ESWI was significantly reduced. We assume $U$ is not a function of $P$, but simulations show that $U$ increases as $P$ decreases (Melinda A. Kahre, via email). This would strengthen the instability.

%Gases with high $\Lambda$ often have a high dipole moment which also makes them highly soluble in water (CO$_2$, NH$_3$, SO$_2$, H$_2$S but not CH$_4$)  and chemically reactive. 
\noindent SDF has very similar requirements to ESWI, but the constraint on rotation rate is stricter. Large-amplitude libration or nonsynchronous rotation would prevent the development of a deep pond around the substellar point. (A low-latitude liquid belt can be imagined, but the idealized EBM of \S2 is not appropriate to that case). Kepler data show that only a small proportion of close-in small-radius exoplanets in multi-planet systems are in mean-motion resonance \citep{lis11}, but that most are close to resonance and could maintain nonzero eccentricity. This would allow for significant nonsynchronous rotation if the planet's spin rate adjusts to keep the substellar point aligned with the star during periapse passage (pseudo-synchronous rotation). In addition, SDF requires very soluble gases: by contrast, the N$_2$ content of even a 100km deep ocean at 298K is only $\sim$0.2 bar per bar of atmospheric N$_2$. In this paper we assume that volatiles are excluded from the ice when the ice freezes. If clathrate phases form, they could absorb volatiles and make the SDF irreversible.

%\emph{Interactions between volcanic degassing, atmospheric pressure, weathering and albedo:}
%\emph{To be added by Eric's group?}

%\subsection{1:1,3:2,2:1? Other spin-orbit resonances}

\subsubsection{Climate evolution into the unstable region}
%\noindent An  arbitrarily small finite perturbation can cause an unstable equilibrium to undergo an instability. Therefore, if a planet has a stable climate, it must have a stable equilibrium. However, 
\noindent Planets could undergo ESWI early in their history if they form in the unstable region. In addition, many common geodynamic and astronomical processes can shift the equilibrium between $W_t$ and $V_n$ (Figure \ref{LSTARP}a), causing a secular drift of the equilibrium across the phase space \{$L_*,\Lambda,V_t$\}  (Figures \ref{WEATHERING} and \ref{HYSTERESIS}). This drift can take a planet from a stable equilibrium to an unstable equilibrium via a saddle-node bifurcation \citep{str94}. % Absence of triggers is unlikely to restrict the range of planets to which the ESWI and SDF could apply.

\noindent$\bullet$ \emph{Dynamics and stellar evolution:} Theory predicts that the secular increase in solar flux should have gradually shifted the position of Earth's climate equilibrium. This is consistent with the sedimentary record of the last 2.5 x 10$^9$ yr \citep{gro93,gro00,rid05,kah07}. For Kepler field (Sunlike, rapidly evolving) stars, this could also occur and potentially cause the ESWI for planets with initially thick atmospheres (Figure 4a).
For M stars, main-sequence insolation changes little over the lifetime of the Universe. 

\noindent$\bullet$ \emph{Atmospheric evolution:} $\Lambda$ can change as atmospheric composition evolves. For example, the rise in atmospheric oxygen following the emergence of oxygenic photosynthesisers probably oxidized atmospheric CH$_4$ and may have caused a catastrophic decline in $\Lambda$ \citep{kop05,dom08}. Even gases with negligible opacity, such as N$_2$, alter $\Lambda$ through pressure broadening \citep{li09}. Carbonate-silicate weathering equilibrium is impossible for planets where atmospheric erosion exceeds geological degassing. For these planets, $V_n$ is negative. Strong stellar winds and high XUV flux are observed for many M stars. Removal of atmosphere by strong stellar winds \citep{mur11} or, for smaller planets, high XUV flux \citep{tia09} could trigger an instability for a planet orbiting an M-star, by reducing $P$ (Figure 4a). %Stability is insensitive to $W_o$, but sensitive to $k_{ACT}$ and the pressure exponent of weathering in Equation 3. For habitable zone rocky planets this trade-off will depend on the details of the deep-time hydrological cycle \citep{pie02}. Impacts eject dust into the atmosphere, but the associated cooling is only likely to be long-lasting enough to trigger instability if the atmosphere is very thin (so that the atmospheric collapse is rapid) or if the ejecta forms an optically thick circumplanetary ring.

\noindent $\bullet$ \emph{Tectonics and volcanism:} Volcanic activity decays with radioactivity \citep{kit09,sle00,sle07,ste03}, so in the absence of tidal heating the equilibrium pressure will gradually fall (on a stable branch where $W_t$ increases with $P$) (Figure 4a). Superimposed on this decline are pulses in volcanic activity due to mantle plumes, and perhaps planetwide volcanic overturns as seem to have occurred on Venus. This will cause spikes in equilibrium pressure. The rate of resurfacing is very sensitive to mantle composition, tidal heating, and tectonic style \citep{kit09,val09,kor10,beh10,beh11,van11}. Shutdown of volcanism (such that $V_n \le$ 0) extinguishes the possibility of a stable climate equilibrium; $P$ will fall monotonically. ESWI can accelerate this decay, and Mars may be an example of this (\S5.1). Mountain range uplift exposes fresh rock and may provide a \emph{O}(10$^7$) yr increase in weathering rate that cools the planet (as arguably and controversially may have occurred for Tibet, Earth: \citet{gar08}). This may trigger ESWI by lowering pressure. More speculatively, drift of continents can cause very large atmospheric pressure fluctuations if greenhouse-gas drawdown occurs mainly on land and is strongly focussed in a high-weathering patch near $\psi$ = 0. A continent drifting over the patch will increase planet-averaged weatherability, and pressure will go down. %A nightside glaciation, which lowers eustatic sea level, can cause a positive feedback if it exposes land near the substellar point. Weathering will increase and draw down atmosphere, $\Delta T_s$ will increase, and expanded nightside glaciation will cause further exposure of land. 

\section{Summary and conclusions}

\noindent Nearby M-dwarfs are targeted by several planet searches:- MEarth \citep{cha09}; the VLT+UVES M-dwarf planet search \citep{zec09}; the VLT+CRIRES M-dwarf planet search \citep{bea10}; HARPS \citep{for11}; M2K \citep{app10}; and proposed space missions TESS \citep{dem09tess}; ELEKTRA; PLATO; and ExoplanetSat \citep{smi10}. These searches are driven in part by the hope that planets orbiting M-dwarves can maintain surface liquid water and be habitable. Maintaining surface liquid water over geological time involves equilibrium between greenhouse-gas supply and removal. Balance is thought to be maintained on habitable planets through temperature-dependent weathering reactions. Climate stability can be undermined by several previously-studied climate instabilities. These include atmospheric collapses \citep{hab94,rea04}, photochemical collapses \citep{zah08,lor97}, greenhouse runaways \citep{kas88,lor99,sug02}, ice-albedo feedback \citep{roe10}, and ocean thermohaline circulation bistability \citep{sto61,epi06}. 
Climate stability can also be undermined if the sign of the dependence of mean surface weathering rate on mean surface temperature is reversed.  This paper identifies two new climate instabilities that involve such a reversal, and are particularly relevant for planets orbiting M-dwarves.

Competition between radiative and advective heat transfer timescales sets surface temperature on synchronously-rotating planets with an atmosphere. The atmosphere moves the surface temperature towards the planetary average, through radiative and turbulent heat exchange, on timescale $t_{dyn}$.  The dayside insolation gradient acts to reestablish gradients in surface temperature, on timescale $t_{rad}$. We refer to planets with $t_{dyn}$ $<$ $t_{rad}$ as dynamically equilibrated, because surface temperature is set by atmospheric dynamics. Venus and Titan are nearby examples. We refer to planets where $t_{dyn}$ $\ge$ $t_{rad}$ as radiatively equilibrated. Mars is a nearby example.

Steeper horizontal temperature gradients promote atmospheric depletion if they stabilize surface liquid films, ponds or oceans in which the atmosphere can dissolve. Once dissolved, the atmospheric gases may be sequestered in the crust by weathering. Weathering rates are much faster when solvents are present and temperatures are high. Weathering and mineral formation can be mediated by thin films of water, and are largely irreversible on habitable-zone planets with stagnant lid geodynamics (karst and oceanic dissolution layers are minor exceptions). Lithospheric recycling may cause metamorphic decomposition of weathering products, returning greenhouse gases to the atmosphere on tectonic timescales. In the absence of weathering, growth of an ocean can reduce atmospheric pressure through dissolution. For example, the fundamental greenhouse gas on Earth is CO$_2$. The partitioning of CO$_2$ between the atmosphere, ocean (solution) and crust (weathering products) is in the ratio 1:50:10$^5$ for Earth \citep{sun07}. Dissolution is fully reversible. %On  locked planets, increasing maximum temperature will spread liquid stability provided that (1) pressure OK (2) (Figure 2). 
Positive feedback occurs if reduced atmospheric pressure further steepens the temperature gradient. Rising maximum temperature resulting from atmosphere drawdown allow further expansion of liquid stability, leading to more drawdown. The zone where liquid is stable spreads over the substellar hemisphere. %, and away from periastron if the orbit is eccentric. 
A halt to the atmospheric collapse occurs when pressure approaches the boiling curve, or when the liquid phase is stable over most of the dayside, or when thermal decomposition by crustal recycling returns weathering products to the atmosphere as fast as they are produced.% [Solubility depends on temperature. Will this brake? Always adding cooler stuff at the margin of the pond: brake only if overall Tpond$\uparrow$ as P$\downarrow$].
%  Figure 2 shows the range of $\Lambda$ and $\chi$ for which the instability may occur. Reducing $\tau$, or reducing $\Lambda$ can move a planet out of the stable field. Raising $T_{av}$ will favor the instability even if $\Lambda$ is held constant. For example, increasing solar luminosity will raise $T_{av}$. For planets with period $p \sim min(t_{dyn},t_{rad})$, increasing eccentricity will raise the effective $T_{av}$.

Our idealized-model results motivate study of the instabilities with GCMs.
 
%Our model predicts a bimodal distribution of day-night temperature contrasts on  locked extrasolar rocky planets. The day-night temperature contrast has been measured for [giants], and will be an early measurement ...

%REEMPHASIZE SENSITIVITIES
%WHAT'S IMPORTANT and WHAT IS NOT

We conclude from this study that:-

\begin{enumerate}
\item Enhanced substellar weathering instability (ESWI) may destabilize climate on some habitable-zone planets. ESWI requires large $\Delta T_s$, which is most likely on planets in synchronous rotation. ESWI does not require strict 1:1 synchronous rotation.
\item Substellar dissolution feedback (SDF) is less likely to destabilize climate. It is only possible for restrictive conditions: small oceans, highly soluble gases, and relatively thin, radiatively weak atmospheres. Furthermore, small amounts of nonsynchronous rotation can eliminate SDF.
%\item
\item The proposed instabilities only work when most of the greenhouse forcing is associated with a weak greenhouse gas that also forms the majority of the atmosphere (it does not work for Earth). There are no exact solar system analogs to ESWI, although Mars comes close. % there are no planets with atmospheres, the  unproven - there are no exact solar system analogs. 
Therefore, it would be incorrect to use these tentative results to argue against prioritizing M-dwarfs for transiting rocky planet searches. 
\item If the ESWI is widespread, we would expect to see a bimodal distribution of day-night temperature contrasts and thermal emission from habitable-zone rocky planets in synchronous rotation. Rocky planets with surface pressures in the unstable region would be rare, so emission temperatures would be either close to isothermal, or close to radiative equilibrium.
\end{enumerate}

\begin{acknowledgements}
\noindent \underline{Acknowledgements.} Itay Halevy collaborated with E.S.K. on the development of this idea for Mars (\S 5.1). We thank Itay Halevy, Ray Pierrehumbert, Rebecca Carey, Dorian Abbott, and especially Ian Eisenman for productive suggestions. Robin Wordsworth and Francois Forget shared output from their exoplanet GCM. E.S.K. thanks Dan Rothman for stoking E.S.K.'s interest in deep time climate stability. E.S.K. and M.M. acknowledge support from NASA grants NNX08AN13G and NNX09AN18G. E.G. is supported by NASA grant NNX10AI90G.
\end{acknowledgements}

\pagebreak

%% the file, uncomment the \plotone, \plottwo, and
%% %\includegraphics commands
%%
%% If you need a layout that cannot be achieved with \plotone or
%% \plottwo, you can invoke the graphicx package directly with the
%% %\includegraphics command or use \plotfiddle. For more information,
%% please see the tutorial on "Using Electronic Art with AASTeX" in the
%% documentation section at the AASTeX Web site,
%% http://www.journals.uchicago.edu/AAS/AASTeX.
%%
%% The examples below also include sample markup for submission of
%% supplemental electronic materials. As always, be sure to check
%% the instructions to authors for the journal you are submitting to
%% for specific submissions guidelines as they vary from

%% This example uses \plotone to include an EPS file scaled to
%% 80% of its natural size with \epsscale. Its caption
%% has been written to indicate that additional figure parts will be
%% available in the electronic journal.

%\begin{figure}
%\epsscale{.80}
%%%\includegraphics{f1}
%\caption{Derived spectra for 3C138 \citep[see][]{heiles03}. Plots for all sources are available
%in the electronic edition of {\it The Astrophysical Journal}.\label{fig1}}
%\end{figure}

\pagebreak

\pagebreak

\begin{figure}[p]
\begin{center}
\includegraphics[width=0.6\textwidth]{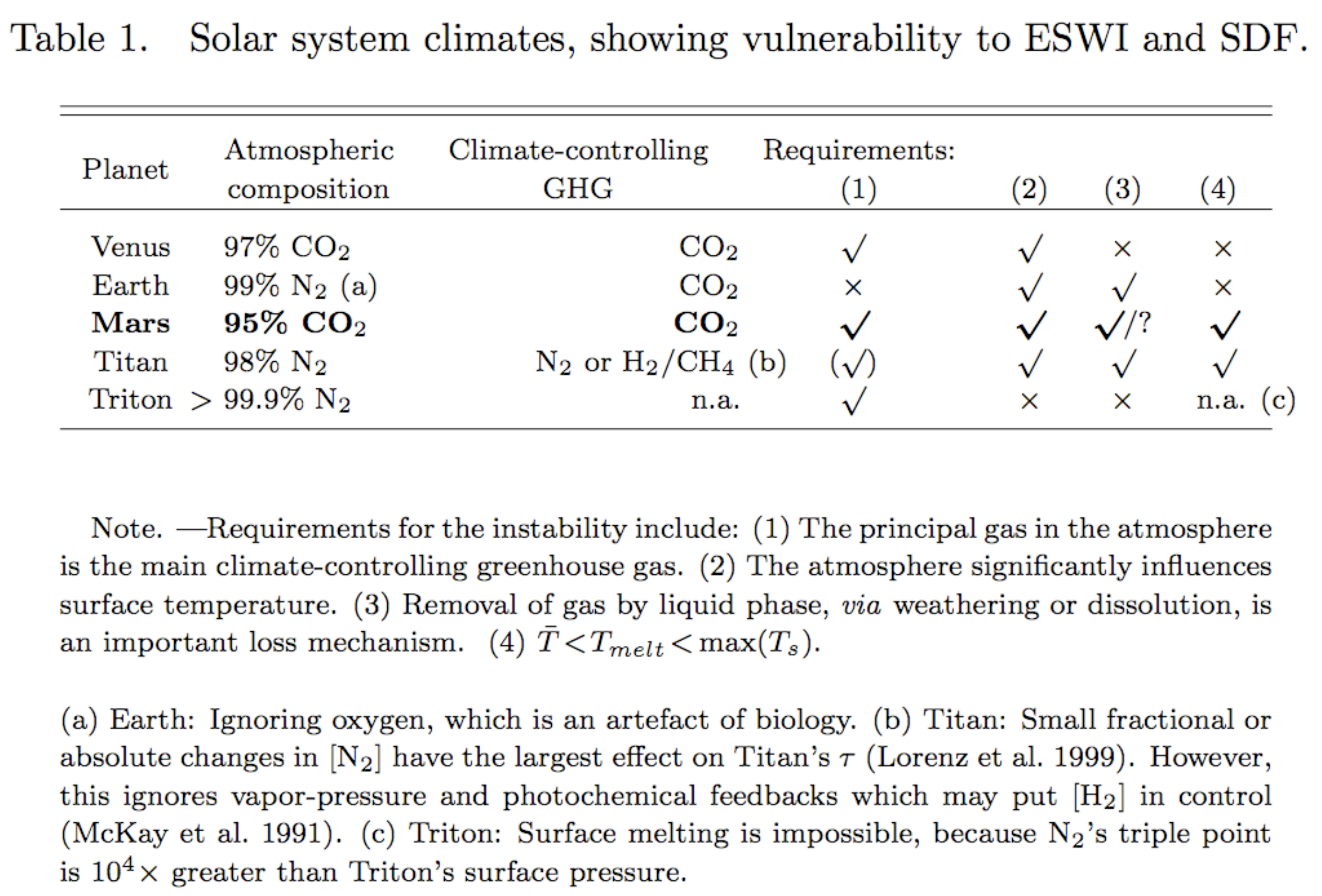}
\end{center}
%\caption{\label{GEOMETRY} Geometry of the idealized energy balance model for an exoplanet in 1:1 spin-orbit resonance. Uneven distribution of starlight ($L_*$) on the planet leads to a hot (white shading, high $T_s$) dayside surface and a cool (black shading, low $T_s$) nightside surface. The atmosphere (uniform gray shading), with horizontally uniform boundary-layer temperature $T_a$, tends to reduce this temperature gradient ($\Delta T_s$). When $T_s > T_{melt}$, a melt pond can form around the substellar point $\psi = 0$, with angular radius $\psi_{max}$ and depth $D_{pond}$. Because rotation is slow, meridional winds are as fast as zonal winds, so $T_s$ depends only on the angular distance from the substellar point ($\psi$).}
% showing coordinates $\psi$ (angular separation from pond), For SDF, show pond radius $\psi_{max}$, $D_{pond}$ ...}
\end{figure}

\begin{figure}[p]
\begin{center}
\includegraphics[width=0.85\textwidth, clip=true, trim = 50mm 120mm 50mm 100mm]{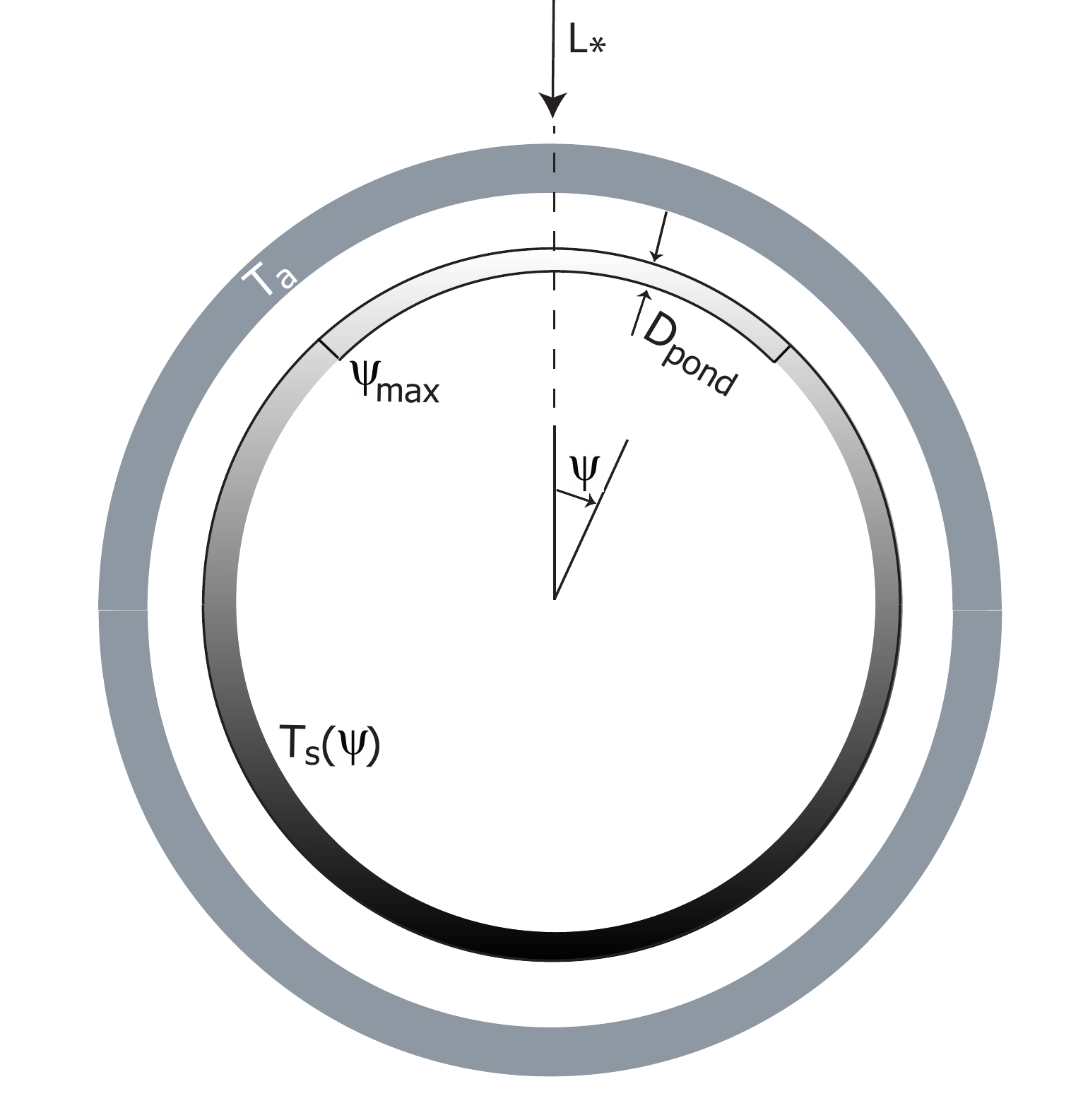}
\end{center}
\caption{\label{GEOMETRY} Geometry of the idealized energy balance model for an exoplanet in 1:1 spin-orbit resonance. Uneven distribution of starlight ($L_*$) on the planet leads to a hot (white shading, high $T_s$) dayside surface and a cool (black shading, low $T_s$) nightside surface. The atmosphere (uniform gray shading), with horizontally uniform boundary-layer temperature $T_a$, tends to reduce this temperature gradient ($\Delta T_s$). When $T_s > T_{melt}$, a melt pond can form around the substellar point $\psi = 0$, with angular radius $\psi_{max}$ and depth $D_{pond}$. Because rotation is slow, meridional winds are as fast as zonal winds, so $T_s$ depends only on the angular distance from the substellar point ($\psi$).}
% showing coordinates $\psi$ (angular separation from pond), For SDF, show pond radius $\psi_{max}$, $D_{pond}$ ...}
\end{figure}

%FIG 6. MAGMA

%FIG 7. SPINORBIT
%Fraction of the radiative-equilibrium day-night surface temperature contrast for surfaces of three thermal inertias: regolith, bedrock, and ice. Insolation = the solar constant. x-axis shows ratio of 
%A thin atmosphere is assumed to keep nightside temperatures above 100K.
%

%\begin{figure}[p]
%\includegraphics[width=1.0\textwidth, clip=true, trim = 0mm 0mm 0mm 0mm]{TEMPFIG.pdf}
%\caption{\label{TSE} Surface temperature as a function of distance from the substellar point in our energy balance model. Diamonds correspond to atmospheric temperature (horizontally uniform).
%In order of increasing temperature, the pressures corresponding to the diamonds are 10$^{-3}$, 10$^{-2}$, 10$^{-1}$, 1 and 10 bars. Arrow is the direction of increasing pressure. $\Lambda$ = 0.1 and $L_*$ = 900 W/m$^2$. }
%\end{figure}

\begin{figure}[p]
\includegraphics[width=1.0\textwidth, clip=true, trim = 25mm 70mm 85mm 180mm]{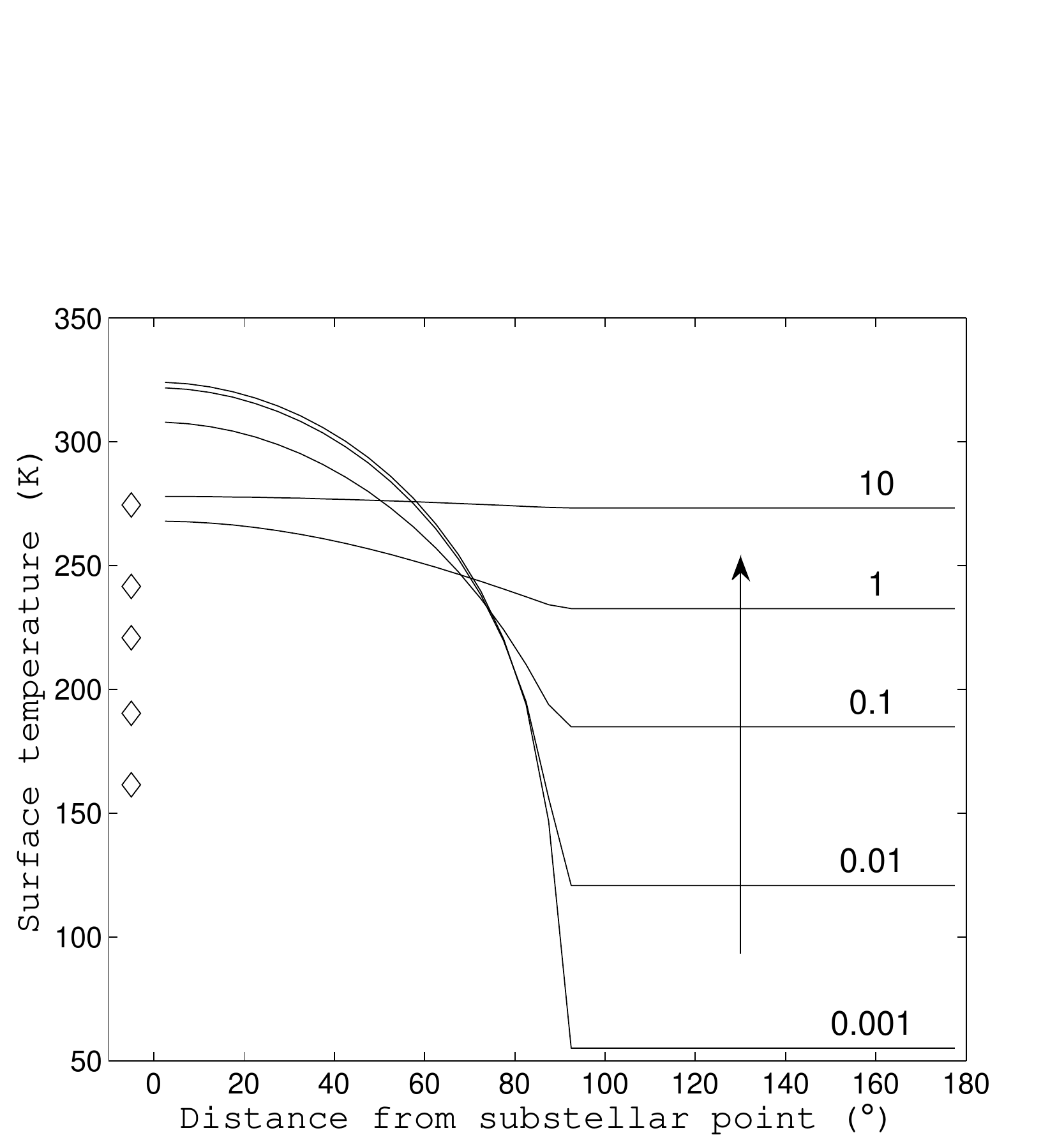}
\caption{\label{TSE} Surface temperature as a function of distance from the substellar point in our energy balance model. Diamonds correspond to atmospheric temperature (horizontally uniform).
In order of increasing temperature, the pressures corresponding to the diamonds are 10$^{-3}$, 10$^{-2}$, 10$^{-1}$, 1 and 10 bars. Arrow is the direction of increasing pressure. Radiative efficiency $\Lambda$ = 0.1, stellar flux $L_*$ = 900 W/m$^2$. }
\end{figure}

 %which are in the same order as the anitstellar temperature 

%\begin{figure}[p]
%\includegraphics[width=1.0\textwidth, clip=true, trim = 0mm 0mm 0mm 0mm]{BIFURCATIONv2.pdf}
%\caption{\label{WEATHERING} Bifurcation diagram to show the enhanced substellar weathering instability for $\Lambda$ = 0.1. 
%The thick black line shows the planet-integrated weathering, $W_t$,  corresponding to the temperature maps shown in Figure \ref{TSE}. If $V_n$ = 0s, $W_t$ $\propto - \left( \frac{\partial P}{\partial t} \right)$. At equilibrium, $\frac{\partial P}{\partial t} = 0$, and $W_t$ equals net supply by other processes, $V_n$. For both the lowest and highest $P$, $W_t \!\! \uparrow$ as $P\!\!\uparrow$. Equilibria on these branches are stable.
%For intermediate pressures, $W_t \!\! \downarrow$ as $P\!\! \uparrow$. The thick dashed line is this unstable branch. The rapid climate transitions which bound the hysteresis loop are shown by vertical arrows. The corresponding unstable equilibria shown by open circles and stable equilibria shown by closed circles.
%The thin black lines correspond to Mars, Earth and Venus insolation
%(in order of increasing normalized weathering rate).
%The shape of the curve is explained in the text. These curves are an 8th-order polynomial fit to the model output. Note that as $L_*\!\!\uparrow$, both inflection points move to higher $P$.}
%\end{figure}

\begin{figure}[p]
\includegraphics[width=1.0\textwidth]{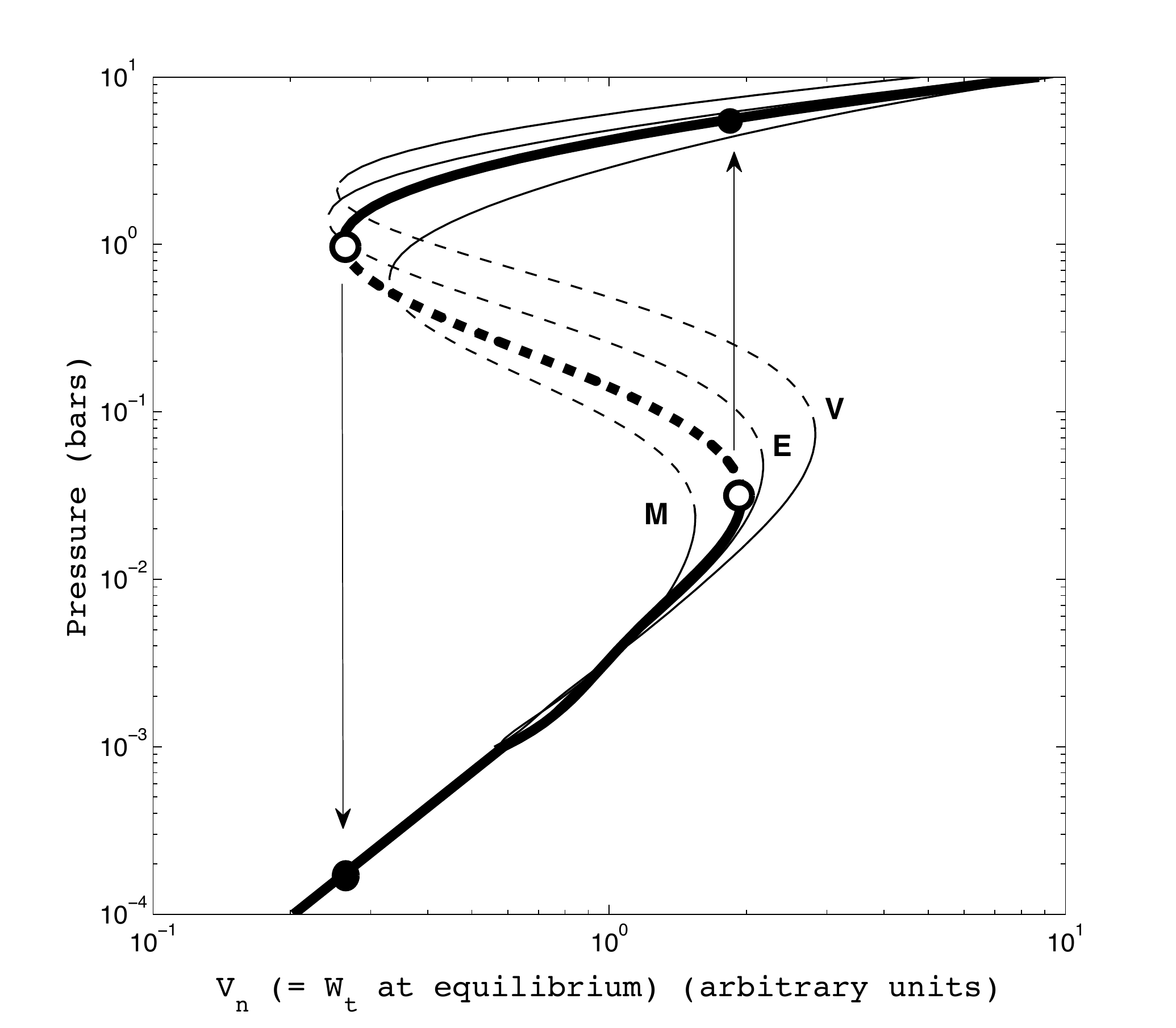}
\caption{\label{WEATHERING} Bifurcation diagram to show the enhanced substellar weathering instability for radiative efficiency $\Lambda$ = 0.1, stellar flux $L_*$ = 900 W/m$^2$. 
The thick black line shows the planet-integrated weathering, $W_t$,  corresponding to the temperature maps shown in Figure \ref{TSE}. If $V_n$ = 0, $W_t$ $\propto - \left( \frac{\partial P}{\partial t} \right)$.  $\frac{\partial P}{\partial t} = 0$ at equilibrium, and $W_t$ equals net supply by other processes, $V_n$. For both the lowest and highest $P$, $W_t \!\! \uparrow$ as $P\!\!\uparrow$. Equilibria on these branches are stable.
For intermediate pressures, $W_t \!\! \downarrow$ as $P\!\! \uparrow$. The thick dashed line is this unstable branch. The rapid climate transitions which bound the hysteresis loop are shown by vertical arrows. The corresponding unstable equilibria are shown by open circles, and stable equilibria are shown by closed circles.
The thin black lines correspond to Mars, Earth and Venus insolation
(in order of increasing normalized weathering rate).
The shape of the curve is explained in the text. These curves are an 8th-order polynomial fit to the model output. Note that as $L_*\!\!\uparrow$, both inflection points move to higher $P$.}
\end{figure}

\vspace{-1.8in}
\begin{figure}[H]
\begin{center}$
\begin{array}{cc}
\includegraphics[width=3.3in]{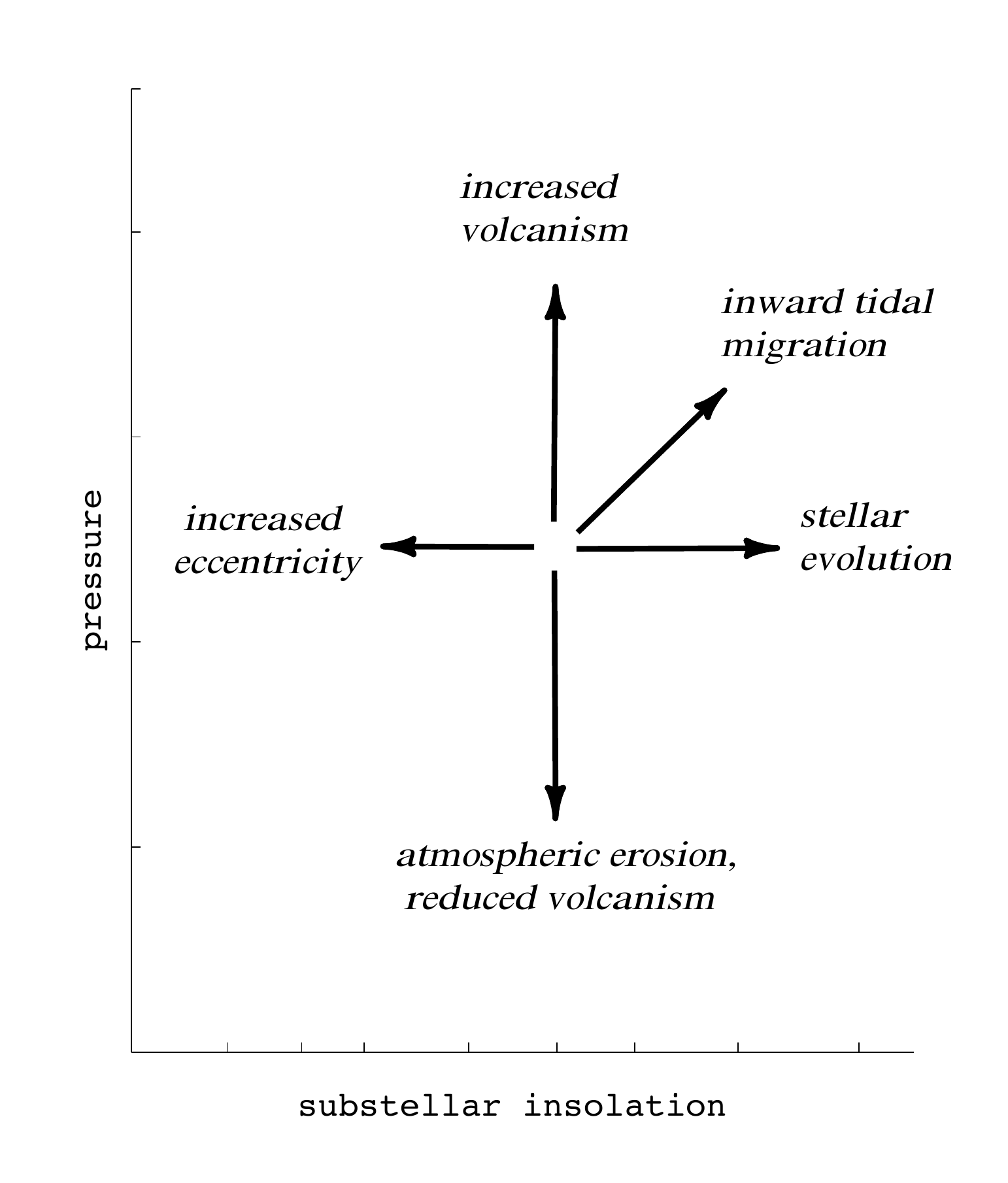} &
\includegraphics[width=3.3in]{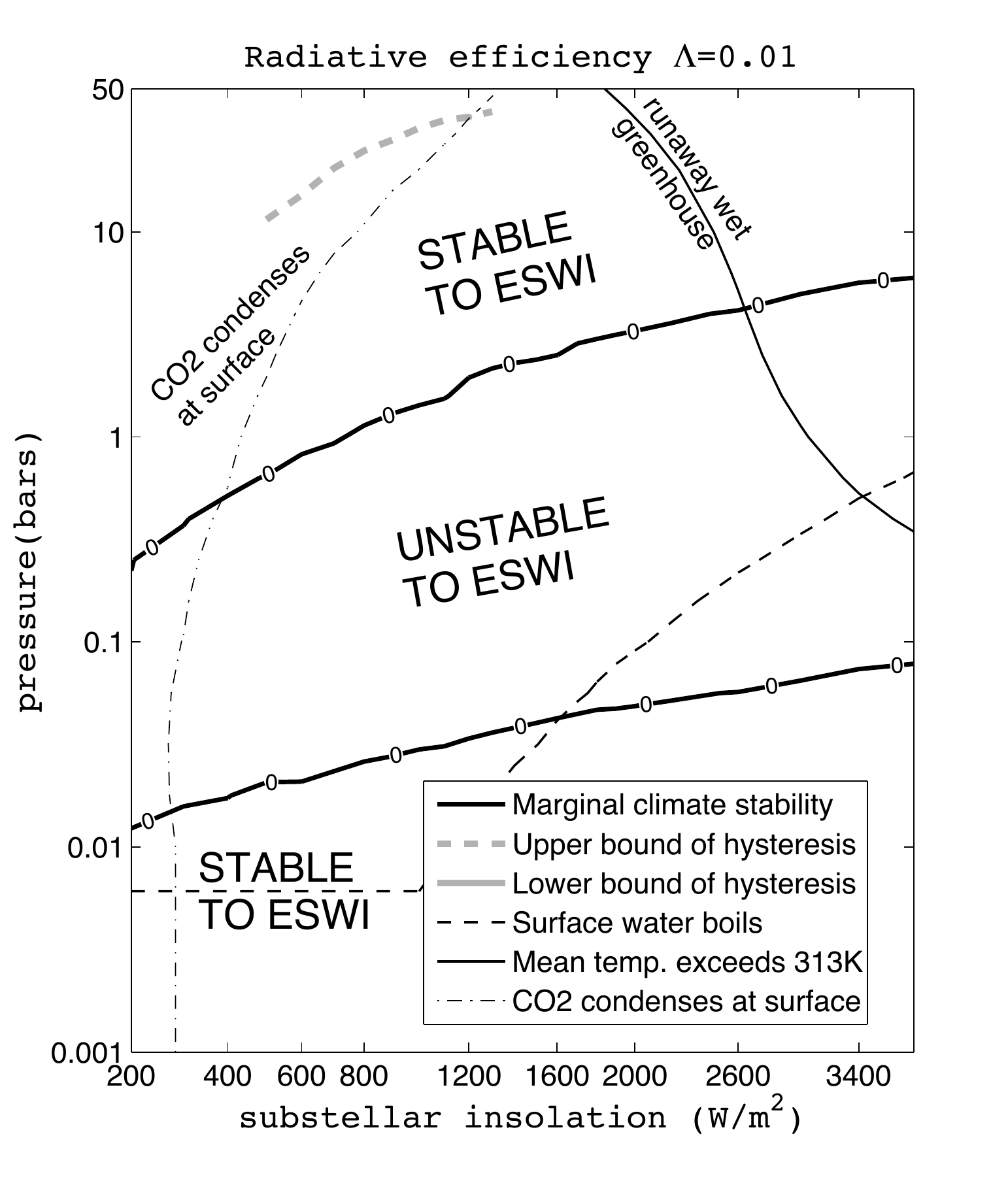}  \\
\includegraphics[width=3.3in]{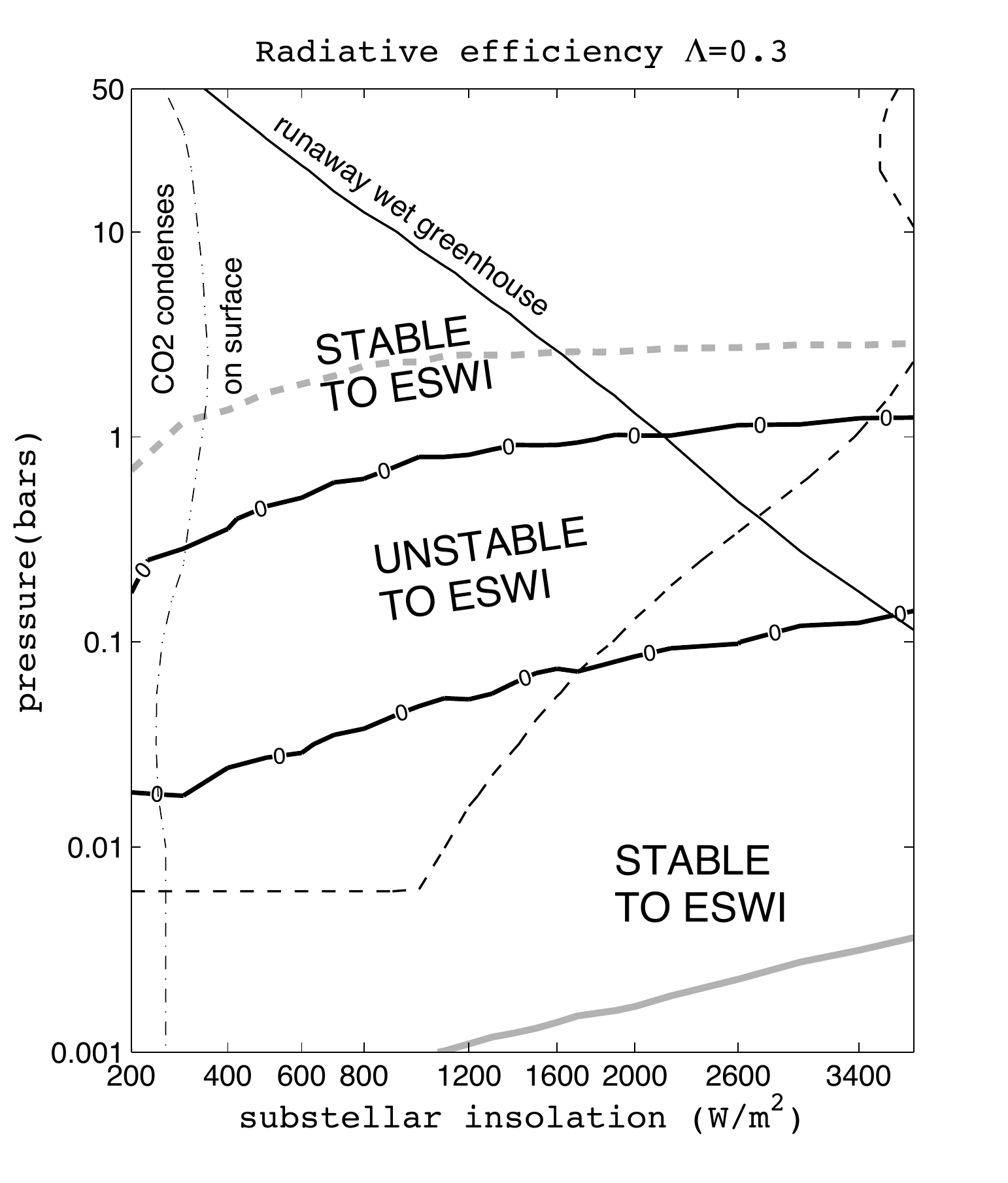} & 
\includegraphics[width=3.3in]{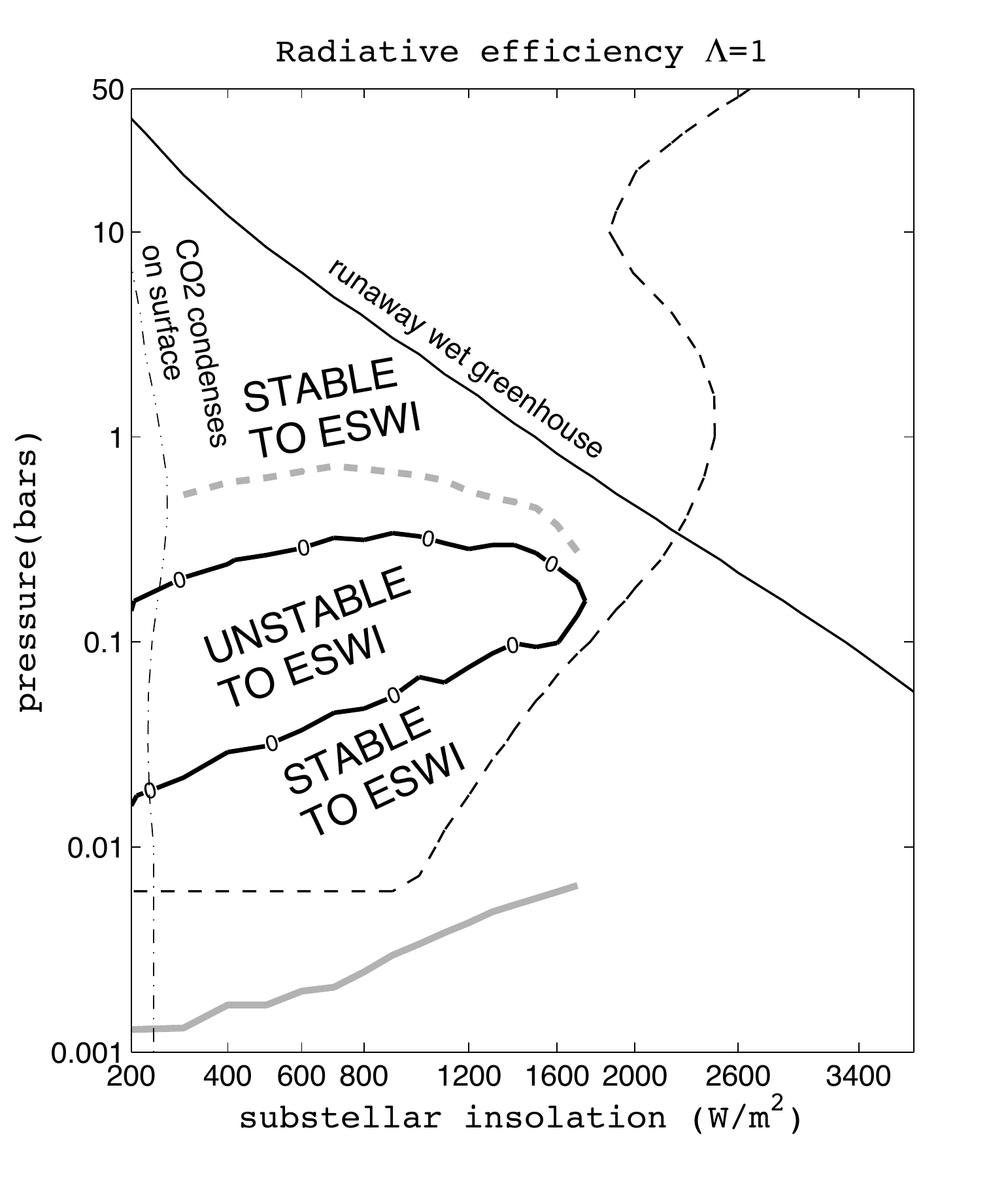} 
\end{array}$
\end{center}
\vspace{-3.7in}
\hspace{3in}
\vspace{15in}
\end{figure}

\newpage
\pagebreak

\begin{figure}
\caption{\label{LSTARP} (a) Mechanisms that can cause secular change in the location of the climate equilibrium $W_t$ = $V_n$. (b,c,d) Habitable zone (HZ) stability diagrams for (b) $\Lambda$ = 0.01, (c) $\Lambda$ = 0.3, (d) $\Lambda$ = 1.0. (Climates with $\Lambda$ $>>$ 1 are always stable against the enhanced substellar weathering instability). The climate states at intermediate pressure within the thick black line labelled with zeros are unstable to ESWI \( \left( \frac{\partial W_t}{\partial P} <0 \right) \). Climates that approach the unstable zone from below will jump up to the dashed gray line. Climates that approach the unstable zone from above will jump down to the solid gray line. These jumps can be extreme; for example, in (b) the solid gray line is everywhere $<$ 0.001 bars (and so is not visible). See the text for discussion of the speed of jumps. The hysteresis loop does not exist for high $L_*$ and and high $\Lambda$, and so the thick gray lines vanish towards the right of (d).  The thin lines correspond to previously-described challenges to habitable-zone climate stability: moist runaway greenhouse (thin solid line); nightside atmospheric condensation of CO$_2$ (dash-dotted line); boiling of surface water (thin dashed line).}
\end{figure}

\begin{figure}[p]
\includegraphics[width=1.0\textwidth]{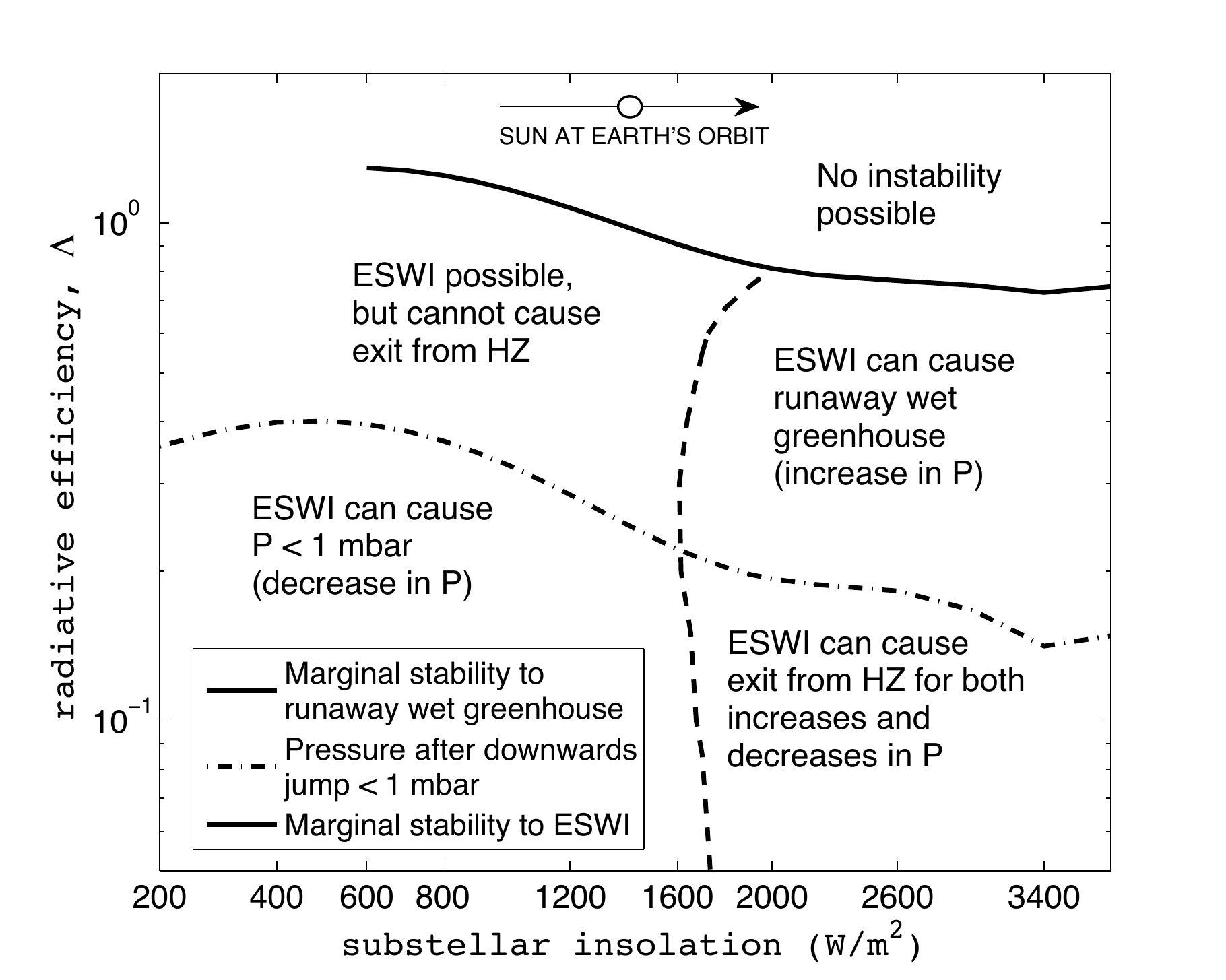}
\caption{\label{HYSTERESIS} Stability phase diagram, showing
the effects of the enhanced substellar weathering instability as a function of $L_*$ and $\Lambda$. The jump in pressure due to ESWI can cause a runaway wet greenhouse (for a jump upwards in $P$), or a decrease in $P$ to the triple point of water (for a jump downwards in $P$). To account for microclimates and solid-state greenhouse effects, we conservatively define ``$P <$ triple point'' as ``$P <$ 1 mbar'', which is below the boiling curves in Figure \ref{LSTARP}. Some curves have been smoothed with a 5th-order polynomial in order to remove small wiggles due to numerical artifacts. The arrow shows the change in stellar flux at 1AU for a solar-mass star over 8 Gyr of stellar evolution in the model of \citet{bah01}, and the circle marks the current solar flux. }  %High radiative efficiency eliminates the instability. The red-line values are the maximum pressures in the unstable branch for which $\left( \frac{\partial W_t}{\partial P} \right) < 0$. The blue-line values are the corresponding minimum pressures. For example, near 650W/m$^2$ and $\Lambda$ $\sim$ 0.2, 3 bars $< P <$ 10 bars cannot be a stable equilibrium. The minimum pressure in the hysteresis loop is shown by blue dotted lines. This is the pressure to which an initially thick atmosphere will collapse if pressure decreases below the red-line values. Frequently these collapses involve $10^3$-fold reductions in pressure. In most cases the top of the hysteresis loop exceeds 50 bars, and we do not show these values. When the top of the hysteresis loop is at $<$50 bars, it is shown by the red dotted line.}%In some cases there is a (red dotted lines).}
%for parameter space \{$L_*, \Lambda$\}
\end{figure}

\begin{figure}[p]
\includegraphics[width=1.0\textwidth, clip=true, trim = 0mm 0mm 0mm 0mm]{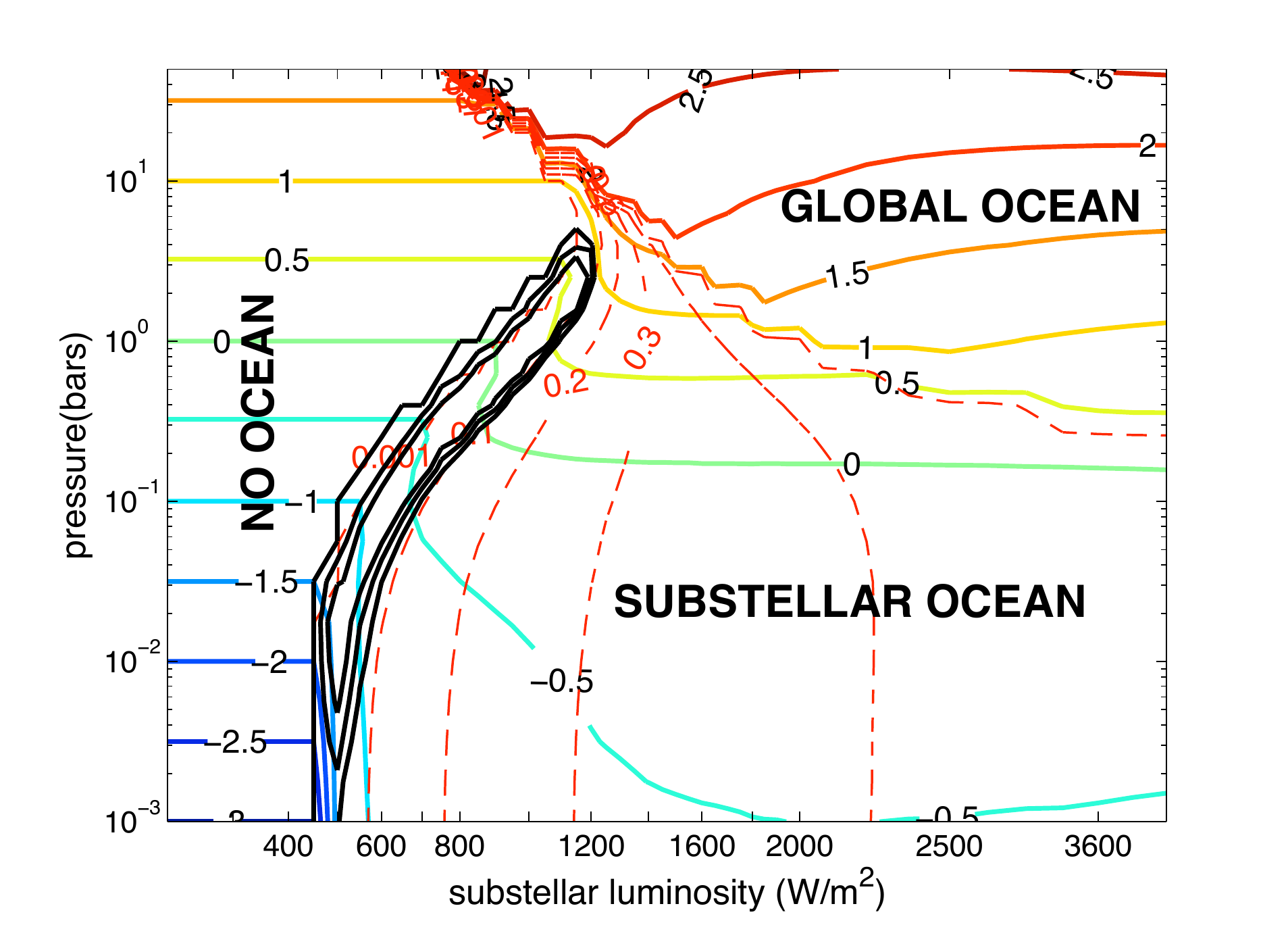}
\caption{\label{SDICO2}Substellar dissolution feedback, for CO$_2$/seawater equilibria, $\Lambda$ = 0.03 (note that $\Lambda$ = 0.03 is unrealistically low for all-CO$_2$ atmospheres), and a 100km-deep ocean. The vertical axis is $P$. Colored solid lines correspond to log10($P$ + $P_{pond}$), i.e. the sum of the atmospheric and ocean inventory. Where these are equal to $P$, there is no ocean. Fractional ocean coverage is shown by the red dashed contours (contour interval is 0.1 in units of planet fractional surface area). Because nightside temperature is constant, fractional ocean coverage jumps from 0.5 (hemispheric ocean) to 1.0 (global ocean). The outermost black line encloses the area where SDF is a positive feedback on small changes in $P$. Outside this area, $\frac{\partial P_{pond}}{\partial P}$ $\ge$ 0 (zero or negative climate feedback). The inner two contours correspond to $\frac{\partial P_{pond} }{\partial P} <$ -0.5 (strong positive feedback) and $\frac{\partial P_{pond} }{\partial P} <$ -1 (runaway). Runaways can only occur for deep oceans and small pond area.} %This is because the slope of the total inventory curve,  $\frac{\partial ( P_{pond} }{\partial P} <$  
%\textbf{Note to draft readers: This is a rather dull result, I expect the same figure for SO$_2$ will be more interesting.} }
\end{figure}


\begin{thebibliography}{}
%\bibitem[Abdulsattar et al.(1977)]{abd77} Abdulsattar, A.H., Sridhar, S., \& Bromley, L.A. 1977, AIChE Journal 23, 62.
{\normalsize
\bibitem[Apps et al.(2010)]{app10} Apps, K., et al., PASP, 122, 156. 
\bibitem[Bahcall et al.(2001)]{bah01} Bahcall, J.~N., 
Pinsonneault, M.~H., \& Basu, S.\ 2001, \apj, 555, 990. 
\bibitem[Bandfield et al.(2003)]{ban03} Bandfield, J.L., Glotch, T.D., \& Christensen, P.R. 2003, Science 301, 1084.
\bibitem[Barabash et al.(2007)]{bar07} Barabash, S., Federov, A., Lundin, R., \& Sauvaud, J.-A. 2007, Science 315, 501.
\bibitem[Batalha et al.(2011)]{kep10b} Batalha, N.M. \& the Kepler Team 2011, ApJ 729, 2011.
\bibitem[Bean et al.(2010)]{bea10} Bean, J.~L., Seifahrt, A., 
Hartman, H., et al.\ 2010, \apj, 713, 410 
\bibitem[B\u{e}hounkov\'{a} et al.(2010)]{beh10} B\u{e}hounkov\'{a}, M., et al. 2010, J. Geophys. Res. - Planets 115(E9), CiteID E09011.
\bibitem[B\u{e}hounkov\'{a} et al.(2011)]{beh11} B\u{e}hounkov\'{a}, M., et al. 2011, ApJ 728, 89.
\bibitem[Berner \& Kothavala(2001)]{ber01} Berner, R.A., \& Z. Kothavala 2001, Am. J. Sci. 301, 182.
\bibitem[Charbonneau et al.(2009)]{cha09} Charbonneau, D., et al. 2009, Nature 462, 891.
\bibitem[Cohen et al.(2004)]{coh04} Cohen, A.S., Coe, A.L., Harding, S.M., \& Schwark, L. Geology 32, 157.
\bibitem[Cowan \& Agol(2011)]{cow11} Cowan, N.B., \& Agol, E. 2011, ApJ 729, 54.
\bibitem[Deming(2009)]{dem09tess} Deming, D. et al. 2009, PASP 121, 952.
\bibitem[Deming \& Seager(2009)]{dem09} Deming, D., \& Seager, S. 2009, Nature 462, 301.
\bibitem[Domagal-Goldman et al.(2008)]{dom08} Domagal-Goldman, S.D., Kasting, J.F., Johnston, D.T., \& Farquhar, J. 2008, Earth Planet. Sci. Lett. 269, 29.
\bibitem[Edmond \& Huh(2003)]{edm03} Edmond, J.M., \& Huh, Y. 2003, Earth Planet. Sci. Lett. 216, 125.
\bibitem[Edson et al.(2011)]{eds11} Edson, A., Lee, S., Bannon, P., Kasting, J.F., \& Pollard, D. 2011, Icarus 212, 1.
\bibitem[EPICA Community Members(2006)]{epi06} EPICA Community Members (first author C. Barbante) 2006, Nature 444, 195.
\bibitem[Exoplanet Community Report(2009)]{exoptf} Lawson, P.R., Traub, W.A. \& Unwin, S.C. 2009, Exoplanet Community Report, Jet Propulsion Laboratory Publication 09--3.
\bibitem[Forveille et al.(2011)]{for11} Forveille, T., Bonfils, X., Lo Curto, G., et al.\ 2011, \aap, 526, A141 
\bibitem[Garzione(2008)]{gar08} Garzione, C.N. 2008, Geology 36, 1003.
\bibitem[Goodwin et al.(2009)]{goo09} Goodwin, P., et al. 2009, Nat. Geosci. 2, 145.
\bibitem[Grotzinger \& James(2000)]{gro00} Grotzinger, J.P., \& James, N.P. 2000, p. 3-20 in Grotzinger, J.P. \& James, N.P. (Eds). Carbonate Sedimentation and Diagenesis in the Evolving Precambrian World, SEPM Sp. Pub. 67, Tulsa, OK.
\bibitem[Grotzinger \& Kasting(1993)]{gro93} Grotzinger, J.P., \& Kasting, J.F. 1993, J. Geol. 101, 235.
\bibitem[Haberle et al.(1994)]{hab94} Haberle, R.M., Tyler, D., McKay, C.P., \& Zahnle, K.J. 1994, Icarus 109, 102.
\bibitem[Halevy et al.(2007)]{hal07} Halevy, I., Zuber, M.T. \& Schrag, D.P. 2007, Science 318, 1903.
\bibitem[Hashimoto \& Abe(2005)]{has05} Hashimoto, G.L., \& Abe Y. 2005, Planet. \& Space Sci. 53, 839.
\bibitem[Hirschmann \& Withers(2008)]{hir08} Hirschmann, M.M., \& Withers, A.C. 2009, Earth Planet. Sci. Lett. 270, 147.
\bibitem[Jackson et al.(2010)]{jac10} Jackson, B., et al. 2010, Mon. Not. R. Astron. Soc. 407, 910.
\bibitem[Jennings et al.(2009)]{jen09} Jennings, D.E. et al. 2009, ApJL 691, L103-L105.
\bibitem[Joshi et al.(1997)]{jos97} Joshi, M.M., Haberle, R.M., \& Reynolds, R.T. 1997, Icarus 129, 450.
\bibitem[Joshi(2003)]{jos03} Joshi, M. 2003, Astrobiology 3, 415.
\bibitem[Kah \& Riding(2007)]{kah07} Kah, L.C., \& Riding, R. 2007, Geology 35, 799.
\bibitem[Kahn(1985)]{kah85} Kahn, R. 1985, Icarus 62, 175.
\bibitem[Kasting(1988)]{kas88} Kasting, J.F. 1988, Icarus 74, 472.
\bibitem[Kasting(1991)]{kas91} Kasting, J.F. 1991, Icarus 94, 1.
\bibitem[Kasting et al.(1993)]{kas93} Kasting, J.F., Whitmire, D.P., \& Reynolds, R.T. 1993, Icarus 101, 108.
\bibitem[Kite et al.(2009)]{kit09} Kite, E.S., Manga, M., \& Gaidos, E. 2009, ApJ 700, 732.
\bibitem[Kite et al.(2011a)]{kit11a} Kite, E.S., Michaels, T.I., Rafkin, S.C.R., Manga, M., \& Dietrich, W.E. 2011a, J. Geophys. Res. 116, E07002, doi:10.1029/2010JE003783.
\bibitem[Kite et al.(2011b)]{kit11b} Kite, E.S., Rafkin, S.C.R., Michaels, T.I., Dietrich, W.E., \& Manga, M. 2011b, J. Geophys. Res., doi:10.1029/2010JE003792, in press.
\bibitem[Knutson et al.(2009)]{knu09} Knutson, H., Charbonneau, D., Cowan, N. B., Fortney, J. J., Showman, A. P., Agol, E., Henry, G.W., Everett, M.E., \& Allen, L.E., ApJ 690, 822.
\bibitem[Kopp et al.(2005)]{kop05} Kopp, R.~E., Kirschvink, 
J.~L., Hilburn, I.~A., 
\& Nash, C.~Z.\ 2005, Proceedings of the National Academy of Science, 1021, 11131. 
\bibitem[Korenaga(2010)]{kor10} Korenaga, J. 2010, ApJL 725, L43.
\bibitem[Kump et al.(2000)]{areps00} Kump, L.R., Brantley, S.L., \& Arthur, M.A. 2000, Ann. Rev. Earth Planet. Sci. 28, 611.
\bibitem[Leeder(1999)]{lee99} Leeder, M.R. 1999, Sedimentology and sedimentary basins: from turbulence to tectonics (1st Edition), Wiley-Blackwell.
\bibitem[L\'{e}ger et al.(2011)]{cor7b} L\'{e}ger, A., \& the CoRoT Team 2009, A\&A 506, 287.
%\bibitem[Lissauer et al.(2011)]{lis11} Lissauer, J.J., et al. 2011, ApJ submitted. 
\bibitem[Lissauer et al.(2011)]{lis11} Lissauer, J.~J., 
Ragozzine, D., Fabrycky, D.~C., et al.\ 2011, arXiv:1102.0543.
\bibitem[Li et al.(2009)]{li09} Li, K.-F., Pahlevan, K., 
Kirschvink, J.~L., 
\& Yung, Y.~L.\ 2009, Proceedings of the National Academy of Science, 106, 9576.
\bibitem[Liu \& Schneider(2011)]{liu11} Liu, J., \& Schneider, T. 2011, J. Atmos Sci., in press.
\bibitem[Lorenz et al.(1997)]{lor97} Lorenz, R.D., McKay, C.P., \& Lunine, J.I. 1997, Science 275, 642.
\bibitem[Lorenz et al.(1999)]{lor99} Lorenz, R.D., McKay, C.P., \& Lunine, J.I. 1999, Plan. and Space Sci. 47, 1503.
\bibitem[Maher(2010)]{mah10} Maher, K. 2010, Earth Planet. Sci. Lett. 294, 101.
\bibitem[Marinova et al.(2005)]{mar05} Marinova, M., McKay, C.P., \& Hashimoto, H. 2005, J. Geophys. Res. 110, E03002, doi:10.1029/2004JE002306.
\bibitem[Matsui \& Abe(1986)]{mat86} Matsui, T. \& Abe, Y. 1986, Nature 319, 303
\bibitem[McKay et al.(1991)]{mck91} McKay, C.P., Pollack, J.B., \& Courtin, R. 1991, Science 253, 1118.
\bibitem[Merlis \& Schneider(2010)]{mer10} Merlis, T.M., \& T. Schneider 2010, Journal of Advances in Modeling Earth Systems 2, 13, doi:10.3894/JAMES.2010.2.13
\bibitem[Mitchell \& Vallis(2010)]{mit10} Mitchell, J.L., \& Vallis, G.K. 2010, J. Geophys. Res. 115, E12008, doi:10.1029/2010JE003587.
\bibitem[Morbidelli et al.(2007)]{nicerecent} Morbidelli, A., Tsiganis, K., Crida, A., Levison, H.F., \& Gomes, R. 2007, AJ 134, 1790.
\bibitem[Mura et al.(2011)]{mur11} Mura, A., et al. 2011, Icarus 211, 1.
\bibitem[Murray \& Dermott(1999)]{mur99} Murray, C.D., \& Dermott, S.F. 1999, Solar System Dynamics, Cambridge.
\bibitem[Murphy et al.(2010)]{mur10} Murphy, B.A., Farley, K.H., \& Zachos, J.C. 2010, Geochem. Cosmochem. Acta 74, 5098.
\bibitem[O'Gorman \& Schneider(2008)]{ogo08} O'Gorman, P.A., \& Schneider, T., 2008, J. Climate 21, 3815.
\bibitem[Phillips et al.(2011)]{phi11} Phillips, R.J., et al. 2011, Science 332, 838. 
\bibitem[Pielke(2002)]{piel02} Pielke, R.A. 2002, Mesoscale Meteorological Modeling, International Geophysics Series 78, Academic Press.
\bibitem[Pierrehumbert(1995)]{pie95} Pierrehumbert, R.T. 1995, J. Atmos. Sci. 52, 1784.
\bibitem[Pierrehumbert(2002)]{pie02} Pierrehumbert, R.T. 2002, Nature 419, 191.
\bibitem[Pierrehumbert(2010)]{pie10} Pierrehumbert, R.T. 2010, Principles of Planetary Climate, Cambridge University Press.
\bibitem[Pierrehumbert(2011)]{pie11} Pierrehumbert, R.T. 2011, ApJL 726, L8.
\bibitem[Pierrehumbert \& Gaidos(2011)]{piegai} Pierrehumbert, R.T., \& Gaidos, E. 2011, ApJL 734, L13.
\bibitem[Pierrehumbert et al. (2011)]{pieareps} Pierrehumbert, R.T., Abbot, D.S, Voigt, A., \& Koll, D. 2011, Ann. Rev. Earth Planet Sci. 39, 417.
\bibitem[Rathbun et al.(2004)]{rat04} Rathbun, J.A., Spencer, J.R., Tamppari, L.K., Martin,T.Z., Barnard, L.  \& Travis, L.D. 2004, Icarus 169, 127.
\bibitem[Read \& Lewis(2004)]{rea04} Read, P.L., \& Lewis, S.R. 2004, The Martian climate revisited: atmosphere and environment of a desert planet, Springer-Praxis.
\bibitem[Richardson \& Mischna(2005)]{ric05} Richardson, M.A., \& Mischna, M.I. 2005, J. Geophys. Res. 110, E03003, doi:10.1029/2004JE002367.
\bibitem[Ridgwell \& Zeebe(2005)]{rid05} Ridgwell, A., \& Zeebe, R.E. 2005, Earth Planet. Sci. Lett. 234, 299.
\bibitem[Roe(2009)]{roe09} Roe, G. 2009, Ann. Rev. Earth Planet. Sci. 37, 93.
\bibitem[Roe \& Baker(2010)]{roe10} Roe, G.H., \& Baker, M.B. 2010, J. Climate 23, 4694.
\bibitem[Sander(1999)]{sancomp} Sander, R. 1999, Compilation of Henry's Law Constants for Inorganic and Organic Species of Potential Importance in Environmental Chemistry (Version 3)
http://www.henrys-law.org
%\bibitem[Scalo et al.(2007)]{sca07} Scalo, J. 2007, Astrobiology 7(1), doi: 10.1089/ast.2006.0125
\bibitem[Schlitzer(2000)]{sch00} Schlitzer, R. 2000, Eos. Trans. AGU 81(5), 45 (\texttt{ewoce.org}).
\bibitem[Schneider et al.(2010)]{sch10} Schneider, T., O'Gorman, P.~A., \& Levine, X.~J.\ 2010, Reviews of Geophysics, 48, 3001 
\bibitem[Segura et al.(2005)]{seg05} Segura, A., et al. 2005, Astrobiology 5(6),Ê706.
\bibitem[Selsis et al.(2011)]{sel11} Selsis, F., Wordsworth, R., \& Forget, F. 2011, A\&A accepted, doi:10.1051/0004-6361/201116654.
\bibitem[Sleep(2000)]{sle00} Sleep, N.H. 2000, \jgr, 105, 17563
\bibitem[Sleep(2007)]{sle07} Sleep, N.H. 2007, ch.9.06 in Treatise on Geophysics, Elsevier.
%\bibitem[Sleep \& Zahnle(2001)]{sle01} Sleep, N.H., \& Zahnle, K. 2001, J. Geophys. Res. 
\bibitem[Sleep \& Zahnle(2001)]{sle01} Sleep, N.~H., \& Zahnle, K.\ 2001, \jgr, 106, 1373. 
\bibitem[Smith et al.(2010)]{smi10} Smith, M.W. 2010, Proc. SPIE 7731, 773127 (2010); doi:10.1117/12.856559
\bibitem[Showman et al.(2010)]{sho10} Showman, A.P., Cho, J. Y-K., \& Menou, K. 2010, Atmospheric Circulation of Exoplanets, in Seager, S. (Ed)., Exoplanets, U. Arizona Press.
%\bibitem[Sobel et al.(2001)]{sob01} Sobel, A.H., Nilsson, J., \& Polvani, L.M. 2001, J. Atmos. Sci. 58, 3650.
\bibitem[Stanley et al.(in press)]{sta11} Stanley, B.D., Hirschmann, M.M., \& Withers, A.C., Geochim. et Cosmochimica Acta, in press, doi:10.1016/j.gca.2011.07.027.
\bibitem[Strogatz et al.(1994)]{str94} Strogatz, S.H. 1994, Nonlinear dynamics and chaos, Perseus Books.
\bibitem[Stevenson(2003)]{ste03} Stevenson, D.J. 2003, Comptes Rendus de l'Acadamie, 335, 99.
\bibitem[Stommel(1961)]{sto61} Stommel, H. 1961, Tellus 13, 224.
\bibitem[Sugiyama et al.(2002)]{sug02} Sugiyama, M., Stone, P.H., \& Emanuel, K.A. 2002, J. Atmos. Sci. 61, 2001.
\bibitem[Sundquist \& Visser(2007)]{sun07} Sundquist, E.T., \& Visser K. 2007, Treatise on Geochemistry, Chapter 8.09, 425-472.
\bibitem[Tarter et al.(2007)]{tar07} Tarter, J.C., Backus, P.R., Mancinelli, R.L., Aurnou, J.M., et al. 2007, Astrobiology 7, 3065.
\bibitem[Taylor et al.(2009)]{tay09} Taylor, L.L., et al. 2009, Geobiology 7, 171.
\bibitem[Tian(2009)]{tia09} Tian, F. 2009, ApJ 703, 905.
\bibitem[Tremain \& Bullock(in press)]{tre11} Tremain, A.H., \& Bullock, M.A., Icarus, in press, doi:10.1016/j.icarus.2011.08.019 
\bibitem[Valencia \& O'Connell(2009)]{val09} Valencia, D., \& O'Connell, R.J. 2009, Earth Planet. Sci. Lett. 286, 492.
\bibitem[Vance et al.(2009)]{van09} Vance, D., Teagle, D.A.H., \& Foster, G.L. 2009, Nature 458, 493.
\bibitem[van Summeren et al.(2011)]{van11} van Summeren, J., Conrad, C.P., \& Gaidos, E., ApJL 736, L15.
\bibitem[Walker et al.(1981)]{hay81} Walker, J.C.G , Hays, P.B., \& Kasting J.F 1981, J. Geophys. Res. 86, 1147.
\bibitem[Werner(2009)]{wer09} Werner, S.C. 2009, Icarus 201, 44.
\bibitem[West et al.(2005)]{wes05} West, A.J., Galy, A., \& Bickle, M. 2005, Earth Planet. Sci. Lett. 235, 211.
\bibitem[White \& Brantley(1995)]{whi95} White, A.F., \& Brantley, S.L. 1995, Chemical weathering rates of silicate minerals, Rev. Mineral., v31, Min. Soc. Am.
\bibitem[Winn et al.(2011)]{55cnce} Winn, J.N., et al. 2011, ApJL 737, L18.
\bibitem[Wordsworth(2011)]{wordsbio} Wordsworth, R.D., arXiv:1106.1411v1 [astro-ph.EP]
\bibitem[Wordsworth et al.(2011)]{wor11} Wordsworth, R.D., Forget, F., Selsis, F., Millour, E., Charnay, B., \& Madeleine, J.-B. 2011, ApJ 733, L48.
\bibitem[Wray et al.(2011)]{wra11} Wray, J.J., et al. 2011, 42nd Lunar and Planetary Science Conference, held March 7Ð11, 2011 at The Woodlands, Texas. LPI Contribution No. 1608, p.2635
\bibitem[Zahnle et al.(2008)]{zah08} Zahnle, K., Haberle, R.M., Catling, D.C., \& Kasting, J.F. 2008, J. Geophys. Res. 113, E11004.
%\bibitem[Zahnle et al.(2010)]{zahnle} Zahnle, K., Schaefer, L., \& Fegley, B. 2010, Cold Spring Harb. Perspect. Biol. 2010;2:a004895.
%\bibitem[Zahnle(2006)]{zahnle} Zahnle, K. 2006, Elements 2(4), 217-222; DOI: 10.2113/gselements.2.4.217
\bibitem[Zechmeister et al.(2009)]{zec09} Zechmeister, M., K{\"u}rster, M., \& Endl, M.\ 2009, \aap, 505, 859. 
\bibitem[Zeebe \& Wolf-Gladrow (2001)]{zee01} Zeebe, R.E., \& Wolf-Gladrow, D. 2001, CO$_2$ in seawater: equilibrium, kinetics, isotopes, Elsevier Oceanography Series, v.65.
\bibitem[Zeebe \& Caldeira(2008)]{zee08} Zeebe, R.E., \& Caldeira, K. 2008, Nature Geoscience 1, 312, doi:10.1038/ngeo18.
\bibitem[Zeebe et al.(2009)]{zee09} Zeebe, R.E, Zachos, J.C. \& Dickens, G.R. 2009, Nature Geoscience 2, doi: 10.1038/ngeo578.}
\end{thebibliography}
\end{document}